\documentclass[journal,comsoc]{IEEEtran}
\usepackage[pass]{geometry}
\usepackage{etex}
\usepackage{subcaption}
\usepackage[font=footnotesize]{caption}
\usepackage{algorithm}
\usepackage{amsfonts}
\usepackage{amsthm}
\usepackage{amsmath}
\usepackage{array}	
\usepackage{bbm}
\usepackage{cite}
\usepackage{dsfont}
\usepackage{epsfig}
\usepackage{float}
\usepackage[T1]{fontenc}
\usepackage{graphicx}

\usepackage{algorithmicx}
\usepackage{algpseudocode}

\newcolumntype{L}[1]{>{\raggedright\let\newline\\\arraybackslash\hspace{0pt}}m{#1}}
\newcolumntype{C}[1]{>{\centering\let\newline\\\arraybackslash\hspace{0pt}}m{#1}}
\newcolumntype{R}[1]{>{\raggedleft\let\newline\\\arraybackslash\hspace{0pt}}m{#1}}

\usepackage{url}
\usepackage{color}
\usepackage{blkarray} 
\usepackage{tikz}
\usetikzlibrary{calc}
\usepackage[export]{adjustbox} 
\usepackage{setspace}

\let\oldFootnote\footnote
\newcommand\nextToken\relax

\renewcommand\footnote[1]{%
    \oldFootnote{#1}\futurelet\nextToken\isFootnote}

\newcommand\isFootnote{%
    \ifx\footnote\nextToken\textsuperscript{,}\fi}

\usepackage{todonotes}


\newtheorem{theorem}{Theorem}{}
\newtheorem{lemma}{Lemma}{}
\newtheorem{corollary}{Corollary}{}
{}

\usepackage[normalem]{ulem} 

\usepackage{booktabs}

\newfont{\mycrnotice}{ptmr8t at 7pt}
\newfont{\myconfname}{ptmri8t at 7pt}

\IEEEoverridecommandlockouts
\allowdisplaybreaks
\begin{document}

\def\sharedaffiliation{%
\end{tabular}
\begin{tabular}{c}}


\title{Stochastic Modeling of Beam Management in mmWave Vehicular Networks }

\author{Somayeh~Aghashahi,~\IEEEmembership{ Student Member,~IEEE,}
Samaneh~Aghashahi,~\IEEEmembership{ Student Member,~IEEE,}
        Zolfa~Zeinalpour-Yazdi,~\IEEEmembership{ Member,~IEEE,}
        Aliakbar~Tadaion,~\IEEEmembership{Senior Member,~IEEE,}
       \\ Arash~Asadi,~\IEEEmembership{ Member,~IEEE}
}
%


\maketitle

\begin{abstract}
Mobility management is a major challenge for the wide-spread deployment of millimeter-wave (mmWave) cellular networks. In particular, directional beamforming in mmWave devices renders high-speed mobility support very complex. This complexity, however, is not limited to system design but also the performance estimation and evaluation. Hence, some have turned their attention to stochastic modeling of mmWave vehicular communication to derive closed-form expressions characterizing the coverage and rate behavior of the network. In this article, we model and analyze the beam management for mmWave vehicular networks. To the best of our knowledge, this is the first work that goes beyond coverage and rate analysis. Specifically, we focus on a multi-lane divided highway scenario in which base stations and vehicles are present on both sides of the highway. In addition to providing analytical expressions for the average number of beam switching and handover events, we provide design insights for the network operators to fine-tune their network within the flexibility provided by the standard in the choice of system parameters, including the number of resources dedicated to channel feedback and beam alignment operations. 

\end{abstract}
\begin{IEEEkeywords}
Millimeter-Wave communication, V2X, 5G NR, Stochastic Geometry
\end{IEEEkeywords}
\IEEEpeerreviewmaketitle

\section{Introduction}

High-throughput connectivity is a necessity for future vehicles to foster support for autonomous driving  and to provide on-board entertainment. The former, in particular, appears to be highly data-intensive due to the large amount of navigational information (e.g., 3D high-resolution maps) which are critical for safe driving and efficient traffic management. The most viable solution to cope with such data-intensive networks is moving towards communication at millimeter-wave (mmWave) frequencies. However, the well-known radio propagation issues at mmWave frequencies require using highly directional beamforming antennas, which increases the complexity of mobility management in highly dynamic networks. 

As a consequence of directional beamforming, mmWave communication requires: $ (i) $ beam alignment between the user and the base station to initialize the connection and $ (ii) $ tracking the user's movement to provide continuous connectivity. There has been significant research on a plethora of different techniques for supporting mobility in mmWave vehicular networks. However, the prevalent techniques for 5G NR became more clear after the first release of 3GPP 5G NR specification, which includes the basic operational details of beamforming in cellular networks.

{\bf Motivation}. The beam management procedure as defined in the standard is a complex process consisting of periodic channel measurement followed by channel state information (CSI) reports as well as non-periodic procedures triggered by external events (e.g., handover). Furthermore, 3GPP foresees a wide configuration range for beam management procedures. For example, the transmission beamforming has a configurable periodicity from $ 5 $~ms to $160 $~ms, which clearly presents a trade-off between system's agility to channel fluctuations and bandwidth occupied by the signaling overhead. Hence, developing methods for computing the achievable network performance under any given standard-defined configuration choice is of paramount importance. In particular, in high mobility vehicular scenarios, the operators require this information to meet the designated level of QoS while minimizing the cost incurred (e.g., number of base stations and operational expenses). 

{\bf Challenge}. Identifying the network performance in highly dynamic scenarios is a formidable task due to the multitude of the variables (e.g., configuration options, network topology, channel fluctuation rate) in play and their interdependencies. In recent years,  many have resorted to stochastic modeling of such large complex networks with the goal of computing the performance as a whole~\cite{Bao:2015vl, ElSawy:2017bn, zeinalpour2018effect}, as opposed to myopic capacity calculation of a small set of nearby cells. However, these works concentrate on cellular networks with static users and without considering the topological particularities native to vehicular networks. Unlike traditional cellular communication, there are only a handful of articles on stochastic geometric modeling of vehicular networks~\cite{cho2018v2x, chetlur2019coverage, sial2019stochastic, wang2018mmwave, yi2019modeling, tassi2017modeling, giordani2018coverage}. 


{\bf Related Work.} The body of work on stochastic modeling of vehicular communication is limited. In~\cite{cho2018v2x, chetlur2019coverage, sial2019stochastic}, the authors derive the coverage probability and the expression of rate under different topologies in sub-6 GHz frequencies. In particular, the authors of~\cite{cho2018v2x} focus on an urban grid topology with 1D Poisson point process (PPP) distribution of vehicles, whereas the works in~\cite{sial2019stochastic, chetlur2019coverage} model the layout of the roads via Poisson line processes (PLP). Although PLPs are more holistic topologically, they do not allow incorporating the impact of number of lanes, and blockages on the road which are essential to mmWave communication. In the above-mentioned papers the conditional  Laplacian of the interference is obtained and the number  and distribution of the distances of the interfering BSs in not investigated. 

The works in~\cite{wang2018mmwave, tassi2017modeling, giordani2018coverage, yi2019modeling} focus on modeling coverage and rate in mmWave vehicular networks. Similar to the above, the authors of~\cite{wang2018mmwave} leverage PLP for modeling the streets, whereas~\cite{yi2019modeling, tassi2017modeling, giordani2018coverage} focus on the multi-lane highway topology. These works made key contributions in characterizing coverage and rate in mmWave vehicular networks. However, to the best of our knowledge, the-state-of-the-art do not characterize the beam switching and handover events and their impact in highly mobile scenarios, which is a key contributor to signaling overhead of vehicular networks.

{\bf Contributions}. Taking the above into account, we model and analyze the effect of beam management and handover in a multi-lane divided highway in which the vehicular users (VUs) and the mmWave BSs are distributed following PPP. To be more realistic, we allow different distributions for BSs on top and the bottom side of the highway. Furthermore, the density of VU process is also non-uniform across different lanes, similar to real road conditions. The following summarizes our contributions:

\begin{itemize}

    \item To the best of our knowledge, this is the first work to provide an analytical and a statistical framework for beam management and handover in mmWave vehicular networks. Prior works~\cite{wang2018mmwave, tassi2017modeling, giordani2018coverage, yi2019modeling} only focused on analysis of coverage and data rate.

    \item We characterize the number of beam switching events by accounting for lateral location of the VU as well as the blockages caused by larger vehicles on the highway. The challenge here lies within deriving an accurate statistical expression for the intricacy of mmWave connectivity. More specifically, the handover and beam switching events not only depends on the location of the BS (Top or bottom of the highway) but also on the sequence and the direction of the handovers (top to bottom, bottom to top, etc). As a result, we first provide the analysis for a highway in which all BSs are deployed at one side (top or bottom). Leveraging the results obtained for the one-side deployment, we complete the analysis of beam switching and handover for a multi-lane divided highway with two-side BS deployment.

    \item We provide detailed evaluation to characterize the impact of beam management on the overall performance of the network. Furthermore, we demonstrate that our detailed model allows the network operators to estimate the network performance under several variables including the density of BSs, vehicular speeds and the beamwidth. 
    
\end{itemize} 

\section{5G NR beam management primer}
\vspace{-1mm}
\label{s:background}

\begin{figure}[t!]
	\centering
	\begin{subfigure}[t]{0.4\textwidth}
	\includegraphics[width=\columnwidth]{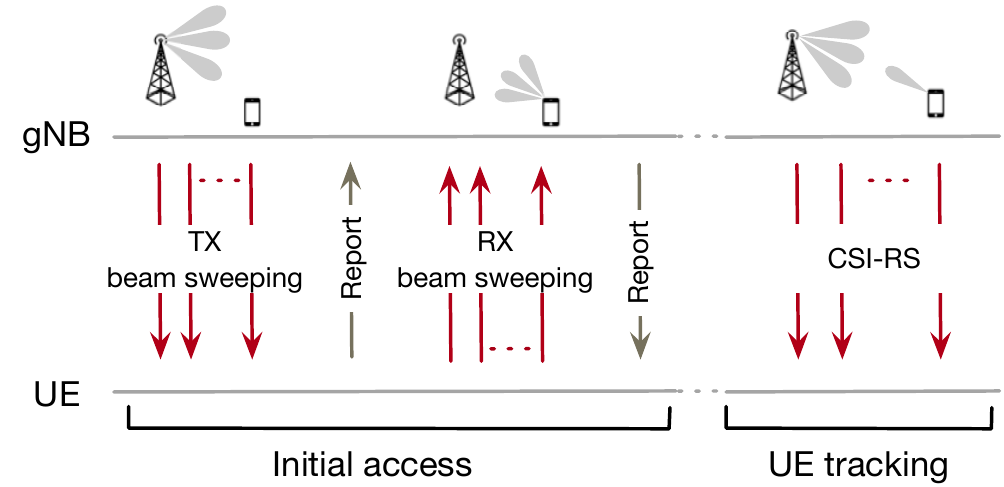}
	\caption{}
	\label{fig:beam_management}
	\end{subfigure}
	
	\begin{subfigure}[t]{0.4\textwidth}
                	\includegraphics[width=\columnwidth]{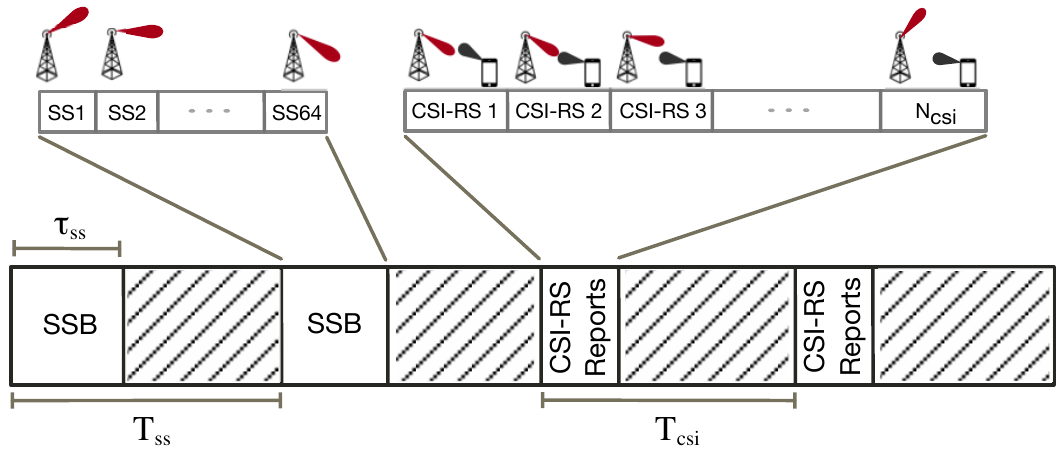}
                	\caption{}
                \label{fig:frame}
	\end{subfigure}
	\caption{A high-level overview of beam management in 5G NR from: (a) protocol perspective, and (b) radio-frame perspective.}
	\label{fig:primer}
	\vspace{-4mm}
\end{figure}

In this section, we provide a high-level overview of the 3GPP standard on beam management as well as the underlying parameters. We encourage the interested reader to refer to a recently published tutorial in~\cite{giordani2018tutorial} or the 3GPP 5G NR specification in~\cite{3GPP38.802, 3GPP38.213,3GPP38.331} for the operational details.

Beam management in 5G NR includes two main procedures:~$ ( i ) $~initial access in which the UE attempts to establish a connection with the gNB through a series of beam sweeping operations, and $ ( ii ) $ UE tracking in which the gNB tunes the beam alignment to adapt to user mobility and changes in the environment.   

%

\subsection{Initial Access}
Given the directionality of the mmWave links, the gNB (base station in 5G terminology) and the UE should align their beams before any communication can take place. Each mmWave device is shipped with a predefined codebook which essentially contains a set of predefined beam patterns. To be able to communicate, the transmitter and the receiver should go through all beam combinations to achieve the desired beam alignment. This procedure is better illustrated in Fig.~\ref{fig:beam_management}. During the initial access phase, first the gNB performs a sweep over all available beams and then the process is repeated at the UE. After each round, a channel report is exchanged to identify the beam which resulted in the highest signal-to-noise ratio (SNR). 

From a radio resource control (RRC) point of view, the initial access can only occur during the synchronization signal burst (SSB) as shown in Fig.~\ref{fig:frame}. The duration of the SSB $\tau_{ss}$ is 5 ms at mmWave frequencies, whereas the periodicity of the SSBs $T_{ss}$ can take any value in $\{5,10,20,40,80,160\}$ ms. Each SSB can only accommodate maximum $ 64 $ SS blocks and only one item of the codebook can be tried in each SS block. For example, if the codebook size of the gNB and UE are $ 48 $ and $12$ respectively, the initial access process will take at least $\frac{48\times12} {64}=9$ SSBs. {\it The trade-off here is between the signaling overhead and the initial access delay.} Lowering $T_{ss}$ results in dedicating a larger part of airtime to the initial access procedure whereas increasing $T_{ss}$ leads to longer waiting time for the UE before it can join the network. 


\begin{figure}[!ht]
    \centering
    \includegraphics*[width=0.9\columnwidth]{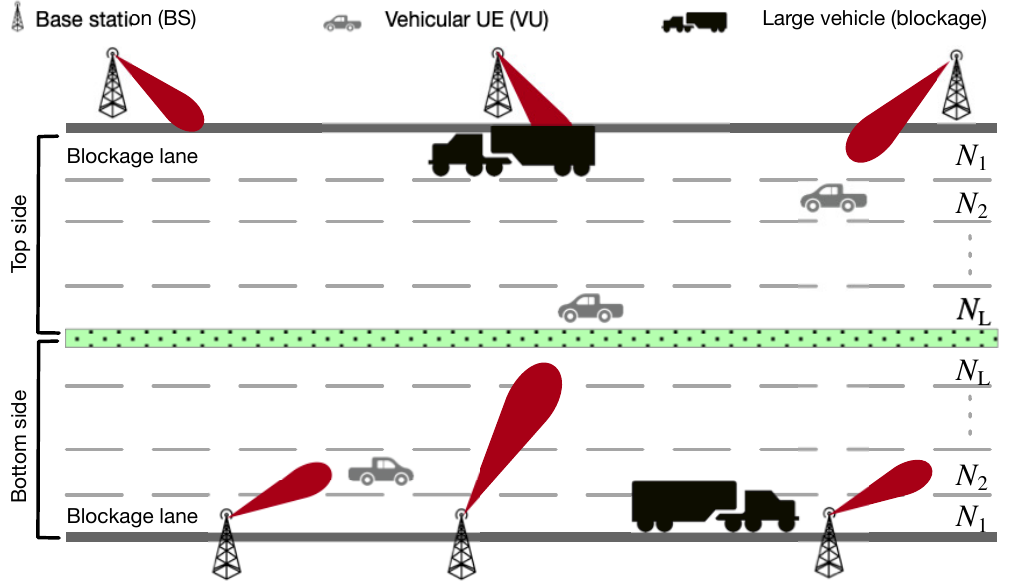}
    \caption{An illustration of the vehicular network under study.}
    \vspace{-4mm}    
    \label{fig:sys_model}
\end{figure}

\subsection{UE Tracking}

The beam alignment achieved during the initial access procedure is subject to failure, e.g., in case of UE mobility. The UE tracking procedure is dedicated to accommodate UE mobility and the challenges associated to it. In 5G NR,  the gNB periodically transmits channel state information reference signal (CSI-RS). More specifically, the gNB sends the CSI-RS on a set of beams, so that the UE can evaluate the quality of the link using other beams. Each CSI-RS can be used for one beam. Note that during CSI-RS, the UE can also monitors different beams. For example, $20$ CSI-RSs are required if the gNB probes $5$ beams from its codebook and the UE listens to these beams in $4$ directions (i.e., $4$ beams from UE's codebook). 

Both the duration of the CSI period $N_{csi}$ and its interval $T_{csi}$ should be decided by the operators. The $N_{csi}$ depends on the number of beams which are monitored at the gNB and the UE, while the periodicity can be $T_{csi} \in \{5, 10, 20, 40, 80, 160, 320, 640\}$ time slots. {\it Note that the choice of  $N_{csi}$ and  $T_{csi}$  determines how fast the network react to user mobility as well as how much overhead the network incurs due to the signaling overhead related to this frequent feedbacks.}

%
%
%

%

\section{System Model}
\label{s:sys_model}

We consider a multi-lane divided highway with $N_L$ lanes in each direction, as shown in Fig.~\ref{fig:sys_model}. We assume that  BSs are deployed at each side of the road following a one-dimensional homogeneous Poisson point process (PPP) with density $\lambda_B$. Let ${\rm BS}_t(i)$         
 denote the $i^{\rm th}$ BS at the top side and ${\rm BS}_b(j)$denote the $j^{\rm th}$ BS at the bottom side of the highway.
The vehicular UEs (VUs) also follow a one-dimensional PPP with density $\lambda_{v,l}$ on the $l^{\rm th}$ lane. The BSs are equipped with antenna arrays and are capable of beamforming under the procedure defined in Section~\ref{s:background}. Each BS has a fixed-size codebook providing beam patterns covering all directions. When the connected VU leaves the coverage area of a beam, either that BS attempts to provide coverage using an alternative beam in the codebook or the VU would be in coverage by another BS on the same or on the other side of the road. 

 Given the innate propagation characteristic of mmWave frequencies, it is important to account for the possible blockages. In a vehicular scenario, the source of blockage is often other larger vehicles (e.g., trucks, buses) on the road. For tractability, we assume that such vehicles only commute on the right most lane ($N_1$ in Fig.~\ref{fig:sys_model}), which is also inline with traffic regulations for such vehicles. We model these blockages as the one-dimensional PPP, whose densities differ at the top side and bottom side of the highway; hence, $\lambda_{t,\mathrm{block}}$  and $\lambda_{b,\mathrm{block}}$ denote the density of large vehicles driving on the right most lane on the top and bottom part of the highway, respectively. 

Given the above blockage model, the probability of having a line-of-sight (LoS) link between a BS and VU is equal to $P_{L,s}= e^{-\tau_{0}\lambda_{s,\mathrm{block}} }, s \in \{t, b\}$, where $\tau_0$ is the length of blockage~\cite{elsawy2013stochastic}.  Leveraging the thinning theorem for PPPs,  the density of the LoS BSs at the top side of the highway for vehicles on each lane will be $\lambda_{s, L}= \lambda_B e^{-\tau_{0}\lambda_{_{s,\mathrm{block}}} }, s \in \{t,b\}$ \cite{baccelli2010stochastic}.  
The pathloss component between the typical VU and an LoS BS with distance $r$ is defined as \cite{andrews2016modeling}:
\begin{equation}
    \ell(r)= C r^{\alpha},
    \label{pathloss_L_N}
\end{equation}
where $\alpha$ is the pathloss exponent and $C$ is the pathloss gain at the unit distance.


\begin{table*}[t!]
\scriptsize
\caption{Table of Notations}
\begin{tabular}{@{}cp{6.1cm} | cp{9cm}@{}}
\textsc{Symbol} & \textsc{Notation} & \textsc{Symbol} & \textsc{Notation} \\
\hline	
\hline
$t$ &         Refers to the top side of the highway.                   &$a_{_k}$ & $\tan (\frac{\pi}{N_{c}}+\frac{2k\pi}{N_{c}})$
\\
 $b$&         Refers to the bottom side of the highway.                     &$\mathrm{N}_s(\mathbf{d}_s) $ & Number of beam switching events, when moving between $\mathrm{BS}_{s}(i)$ and $\mathrm{BS}_{s}(i+1)$. \\
$s$  &          $s \in\{t,b\}$                 &$\mathrm{N}_{tb} $  & Number of beam switching events, when moving between consecutive BSs, $\mathrm{BS}_{t}(i)$ and $\mathrm{BS}_{b}(j)$.\\
$\mathrm{BS}_{s}(i)$&The $i^{\rm th}$ BS at side $s$ of the highway. &$\mathrm{N}_{bt}$& Number of beam switching events, when moving between consecutive BSs, $\mathrm{BS}_{b}(j)$ and $\mathrm{BS}_{t}(i+1)$.\\
$\mathbf{b}_{s,i}$& The x-axis coordinate of $\mathrm{BS}_{s}(i)$. &$\mathbf{h}_{s,s^{\prime}}$& The handover point between consecutive BSs, $\mathrm{BS}_{s}(i)$ and $\mathrm{BS}_{s^{\prime}}(j)$.  \\
$L_h$ &         The length of the highway.                &$\mathrm{N}_{th} $   &Number of beam switching events, when moving between $\mathbf{b}_{t,i}$ and $\mathbf{h}_{t,b}$.  \\
 $N_L$&  			Number of the lanes  in each direction of the highway.				&$\mathrm{N}_{hb} $ &  Number of beam switching events, when moving between  $\mathbf{h}_{t,b}$ and $\mathbf{b}_{b,j}$.\\
$\lambda_{v,l}$ &  Density of the vehicles on the $l^{\rm th}$ lane.		&$\mathrm{BSN}_s$ & Average number of beam switching events when  BSs deployed on side $s$.\\
$\lambda_{s,\mathrm{block}}$ &  Density of blockages on the right most lane of side $s$.    &$\mathrm{BSN}$  &Average number of beam switching events with double-side BS deployment. \\
$\tau_{0}$ & 			The length of the	vehicular	 blockages.					 &$\mathrm{HON}_s$&Average number of handover events along the highway with  BSs deployed on side $s$. \\
 $C$ &  			The pathloss	gain.									& $\mathrm{HON}$ &Average number of handover events along the highway with double side deployed BSs. \\
  $\alpha$ & The pathloss	exponent.	 & $\mathrm{NS}_{t,\mathrm{box}}$&Average number of beam switching events between $\mathrm{BS}_{t}(i)$ and $\mathrm{BS}_{t}(i+1)$.  \\
$\lambda_B$  & The density of the BSs at each side of the highway. & $\mathrm{NH}_{t,\mathrm{box}}$ & The average number of handover events between $\mathrm{BS}_{t}(i)$ and $\mathrm{BS}_{t}(i+1)$.   \\
$\lambda_{s,L}$ & The density of the LoS BSs at  side $s$ of the highway. &$P_{ss^{\prime}}$ & The probability of handover between $\mathrm{BS}_{s}(i)$ and $\mathrm{BS}_{s^\prime}(j)$.  \\
$\lambda_{t,b,L}$ & $\lambda_{t,L}+\lambda_{b,L}$  &$P_{ss}$ &The probability of handover between $\mathrm{BS}_{s}(i)$ and $\mathrm{BS}_{s}(i+1)$. \\
  $N_c$& The size of the codebook at each BS.   & $\mathbf{n}_b$.& Number of the bottom side BSs between $\mathrm{BS}_{t}(i)$ and $\mathrm{BS}_{t}(i+1)$. \\
    $\mathbf{d}_s$ & The distance between $\mathrm{BS}_{s}(i)$ and $\mathrm{BS}_{s}(i+1)$.      &$\mathbf{n}_{bv}$. &Number of the bottom side BSs between $\mathrm{BS}_{t}(i)$ and $\mathrm{BS}_{t}(i+1)$, which serve the VU.
   \\
 $W_{s,l}$ & The distance between the BS antenna and center line of the $l^{\rm th}$ lane. & &  \\ 
\bottomrule 
\end{tabular}
\label{tab:notations}
\vspace{-5mm}
\end{table*}

As mentioned, the BS  is capable of beamforming. We assume that each BS has an orthogonal and a symmetrical codebook with size $N$. The orthogonality aims at reducing interference between beam patterns defined within the codebook. As shown in Fig.~\ref{fig:codebook}, the orthogonality is defined with reference to 3dB beamwidth and there is still interference between beams as discussed in Section~\ref{s:sys_model}. Assuming a circular shape around the antenna array (shown with blue-dotted circle in the figure), the beam patterns in a symmetric codebook are designed such that the quarters of the circle are symmetrical (see Fig.~\ref{fig:codebook}). The symmetricity ensures that our analysis is independent of the  orientation of the antenna. Note that we can derive the number of beam switching events for a non-symmetrical codebook with a similar calculations, however, for brevity we did not include those in this article. 

Also to maximize link quality, we assume that the users associate to the BS which provides the best SINR. In our network, such an association policy entails in choosing the LoS BS which incurs the lowest pathloss (i.e., the nearest BS). 
In the following,  we first obtain the average number of beam switching and handover events for a highway with single-side deployment of the BSs in Section~\ref{ss:single_side}. Next, we extend our analysis to a double-side deployment scenario in Section~\ref{ss:divided}.

\textbf {Notations}: In this paper, we use bold face to refer to the random variables and regular font for the deterministic parameters. Also $\mathbb {E}\{\mathbf{x}\}$ denote the expectation value of random variable $\mathbf{x}$ and $u(.)$ is the unit step function. Table \ref{tab:notations} reviews the definitions of the random variables and also the network parameters.


\begin{figure}[!t] 
    \begin{subfigure}{.16\textwidth}
        \includegraphics*[width=1\columnwidth]{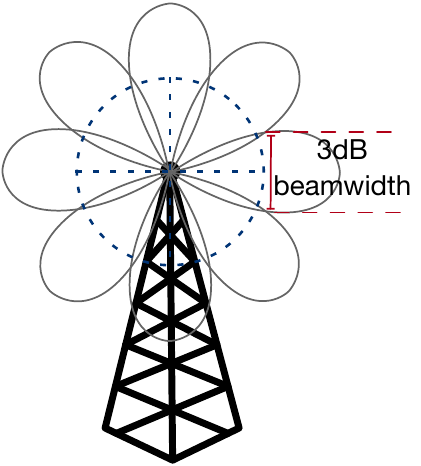}
        \caption{}
        \label{fig:codebook}
    \end{subfigure}
\hspace{-2mm}    
    \begin{subfigure}{.36\textwidth}
        \includegraphics*[width=1\columnwidth]{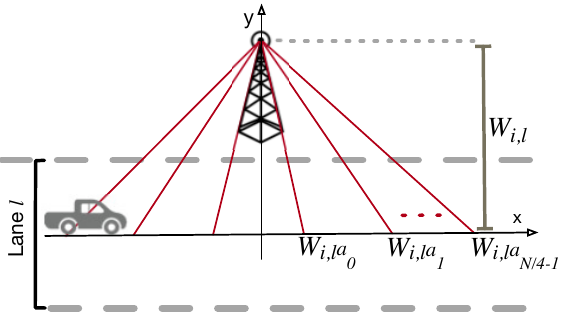}
        \caption{}
        \label{fig:beamswitching}
    \end{subfigure}
    \caption{Illustration of the codebook and its mapping on the road.}
    \vspace{-4mm}    
\end{figure}

\section{Beam Switching and Handover Events Analysis for the Highway with single-side BS deployment}
\label{ss:single_side}
In this section, we derive the analytical expressions for the average number of beam switching and handover events in the above-described vehicular network where BSs are deployed on one side of a multi-lane divided highway. 
Under this 
assumption, in Lemma~\ref{lem:num_4}, we calculate the number of beam switching events for a typical VU (located at the $l^{\rm th}$ lane), which traverses between  two consecutive BSs, ${\rm BS}_s (i)$ and ${\rm BS}_s (i+1)$, conditioned on the distance between these two BSs, i.e. $\mathbf{d}_s$, where $s\in\{t,b\}$ and then average the result over the distribution of $\mathbf{d}_s$ in  Theorem~\ref{the:beam_switch1}. Finally we extend the result to the entire highway in Corollary~\ref{cor:avg_beam_high}.

Assuming all BSs are at side $s$, $s \in \{t,b\}$, where $t$ refers to  the top and $b$ refers to the bottom side of the road, the distance between the serving BS antenna and a typical VU at the $l^{\rm th}$ lane is denoted by $W_{s,l}$, see Fig.~\ref{fig:beamswitching}. 
We must note that the symmetricity assumption forces the codebook size to be a multiple of 4, i.e., $N_{c}=4K, \ K \in \mathbb{N}$.


\begin{lemma}
\label{lem:num_4}
 Assuming that the distance between the serving BS and the next BS (both on side $s \in \{ t,b\}$ of the highway) is $\mathbf{d}_{s}$, the handover occurs when the VU is located at equal distance from both BSs. Consequently, the number of beam switching events, when moving between two neighboring BSs with distance $\mathbf{d}_{s}$, can be expressed as

\begin{equation}
    \label{num_4} 
    \mathrm{N}_{s}(\mathbf{d}_{s})= 
    \begin{cases}
        0  & 0<\mathbf{d}_{s}< 2 W_{s,l}a_{_{0}} 
        \\
        2 & 2 W_{s,l}a_{_{0}}  <\mathbf{d}_{s}< 2 W_{s,l} a_{_{1}}
        \\
        \vdots
        \\
        2k & 2 W_{s,l} a_{_{k-1}}  <\mathbf{d}_{s}< 2 W_{s,l} a_{_{k}}
        \\
        \vdots 
        \\
        \frac{N_{c}}{2} & 2W_{s,l} a_{_{\frac{N_{c}}{4}-1} } <\mathbf{d}_{s}
    \end{cases},
\end{equation}
where $a_{_k}=\tan (\frac{\pi}{N_{c}}+\frac{2k\pi}{N_{c}}) $.
\begin{IEEEproof}
The association policy is minimum pathloss component; thus the handover between the two neighboring BSs happens when the VU is located at equal distance from both BSs, such as the projection of the VU to the BSs lane is at the middle of the line connecting the serving BS and the next BS (i.e., VU fall on the perpendicular bisector of the line connecting the two serving BSs). Hence the distance between each BS and the projection of the VU to the BSs lane is $\frac{\mathbf{d}_s}{2}$. 

As shown in Fig.~\ref{fig:beamswitching}, we place the $x-\text{axis}$ at the middle of the $l^{\rm th}$ lane and set the projection of the serving BS on this lane as the point of origin (0,0).
 Assuming $N_{c}=4K$, the beams intersect with the half positive of the $x-\text{axis}$ at $\frac{N_{c}}{4}$ points, on $ W_{s,l} \tan \frac{\pi}{N_{c}} $,  $ W_{s,l} \tan \frac{3\pi}{N_{c}}$,$\dots$,  $W_{s,l}\tan (\frac{\pi}{N_{c}}+\frac{2(\frac{N_{c}}{4}-1)\pi}{N_{c}})$, as shown in Fig.\ref{fig:beamswitching}).
 Thus, the number of beam switching events under the serving BS and prior to the handover point can be expressed as a function of $\mathbf{d}_{s}$ as follows

 \begin{eqnarray}
 \mathrm{Num}(\mathbf{d}_{s})=
      \begin{cases} 
            0  & 0<\mathbf{d}_{s}< 2 W_{s,l} a_{_{0}} 
            \\
            1 & 2 W_{s,l} a_{_{0}}  <\mathbf{d}_{s}< 2 W_{s,l} a_{_{1}}
            \\
            \vdots
            \\
            k & 2 W_{s,l} a_{_{k-1}}  <\mathbf{d}_{s}< 2 W_{s,l} a_{_{k}}
            \\
            \vdots 
            \\
            \frac{N_{c}}{4} & 2 W_{s,l} a_{_{\frac{N_{c}}{4}-1} } <\mathbf{d}_{s}
    \end{cases},
 \end{eqnarray}
and because of the symmetricity of the scenario between two neighboring BSs, the number of beam switching events from the current serving BS to the next serving BS at distance $\mathbf{d}_{s}$ ($\mathrm{N}_{s}(\mathbf{d}_{s})$) is    equal to  $2\mathrm{Num}(\mathbf{d}_{s})$ which is presented in the Theorem.
\end{IEEEproof}
\end{lemma}

Next, we relax the conditional distance $\mathbf{d}_{s}$ between two serving BSs and compute the average number of beam switching events between two neighboring BSs via the following theorem.

\begin{theorem}
\label{the:beam_switch1}
Assuming that all BSs are located at side $i$ of the highway, the average number of beam switching events while the typical VU at the $l^{\rm th}$ traverses from the serving BS to the next BS within the coverage area\footnote{We define the coverage area between two BSs as the portion of the highway whose limits are the projection of these BSs on the x-axis.}  is 
\begin{align*} 
\mathbb{E}\{\mathrm{N}_{s}(\mathbf{d}_{s})\}=2\sum_{k=1}^{\frac{N_{c}}{4}}\exp\{-2W_{s,l} \lambda_{s,L} a_{_{k-1}}\}.
\end{align*}
\begin{IEEEproof}
 Computing the expectation over the beam switching events, we have:
\begin{align*}
& \mathbb{E}\{ \mathrm{N}_{s}(\mathbf{d}_{s})\}=2\int_{x} \mathrm{N}_{s}(x) f_{_{\mathbf{d}_{s}}}(x) d x =2\int_{2W_{s,l} a_{_0}}^{ 2 W_{s,l} a_{_1} }  f_{_{\mathbf{d}_{s}}}(x) d x+\dots\\ &+2\int _{2W_{s,l} a_{_{\frac{N_{c}}{4}-2}}}^{ 2W_{s,l} a_{_{\frac{N_{c}}{4}-1}} }(\dfrac{N_{c}}{4} -1)f_{_{\mathbf{d}_{s}}}(x) d x  +2\int_ { 2W_{s,l} a_{_{\frac{N_{c}}{4}-1}}}^{\infty} \dfrac{N_{c}}{4} f_{_{\mathbf{d}_{s}}}(x) d x
\end{align*}
where $ f_{_{\mathbf{d}_{s}}}(.)$ is the probability density function of the distance between two consecutive points of a PPP with density $\lambda_{s,L}$, and thus
$f_{\mathbf{d}_{s}}(x)=\lambda_{s,L}e^{-\lambda_{s,L}x} u(x).$
Hence, the expectation of $\mathrm{N}_{s}(\mathbf{d}_{s})$ can be expressed as:
\begin{align*}
& \mathbb{E}\{ \mathrm{N}_{s}(\mathbf{d}_{s})\}=2 \sum_{k=1}^{\frac{N_{c}}{4}-2} k \int_{ 2 W_{s,l} a_{_{k-1}}}^{ 2 W_{s,l} a_{_k} }\lambda_{s,L}e^{-\lambda_{s,L}x} d x\\ &+\dfrac{N_{c}}{2}\int_ { 2W_{s,l} a_{_{\frac{N_{c}}{4}-1}}}^{\infty}  \lambda_{s,L}e^{-\lambda_{s,L}x} d x \\  & =
2\sum_{k=1}^{\frac{N_{c}}{4}-2} k \Big(\exp\{ -2 W_{s,l} a_{_{k-1}})\}-2\exp\{ -2 W_{s,l} a_{_k} \}\Big)+\\ & \dfrac{N_{c}}{2}\exp\{ -2W_{s,l} a_{_{\frac{N_{c}}{4}-1}}\}=2\sum_{k=1}^{\frac{N_{c}}{4}}\exp\{-2W_{s,l} \lambda_{s,L} a_{_{k-1}}\}
\end{align*}
\end{IEEEproof}
\end{theorem}

Knowing the average beam switching events between two neighboring BSs, we can extend the above result to the entire highway with the following corollary. 

\begin{corollary}
\label{cor:avg_beam_high}
The average number of beam switching events along the highway can be expressed as

\begin{align*}
& \mathrm{BSN}_{s}=2\lambda_{s,L}L_h
\sum\limits_{k=1}^{\frac{N_{c}}{4}}\exp\{-2W_{s,l} \lambda_{s,L} a_{_{k-1}}\},
\end{align*}
where $L_h$ is the length of the highway and $a_{_k}=\tan(\frac{\pi}{N_{c}}+\frac{2k\pi}{N_{c}})$. In addition the average number of the handovers is equal to

\begin{equation*}
\mathrm{HON}_{s}=\lambda_{s,L}L_h-1 
\end{equation*}

\begin{IEEEproof}
The average number of beam switching events along the highway can be obtained by averaging the summation of the beam switchings under each serving BS. Considering  $\mathbf{n}_{B,s}\sim \mathrm{poiss}(L_h \lambda_{s,L}), s \in \{t,b\}$ as the number of LoS BSs along each side of the highway, the average number of  beam switching events along the highway would be
\begin{align*}
\mathrm{BSN}_{s} &= \mathbb{E} \{\mathbf{n}_{B,s} \mathrm{N}_{s}(\mathbf{d}_{s})\}= \mathbb{E} \{\mathbf{n}_{B,s}\} \mathbb{E} \{\mathrm{N}_{s}(\mathbf{d}_{s})\}\\ &=2L_h \lambda_{s,L}\sum\limits_{k=1}^{\frac{N_{c}}{4}}\exp\{-2W_{s,l} \lambda_{s,L} a_{_{k-1}}\},
\end{align*}
where the last equality is obtained from Theorem \ref{the:beam_switch1}.
Note that while moving along the highway, the VU can only connect to the LoS BSs. Hence, the average number of handovers is equal to the average number of LoS BSs minus one
\begin{align*}
\mathrm{HON}_{s}=\mathbb{E}\{\mathbf{n}_{B,s}-1\}=L_h \lambda_{s,L}-1
\end{align*}
\end{IEEEproof}
\end{corollary}

\section{Beam Switching and Handover Events Analysis for the Highway with double-side BS deployment}
\label{ss:divided}

In this section, we assume that BSs are distributed at both sides of the highway and VU can be served by BSs on either side of the highway. We obtain the average number of beam switching events in this scenario.
Note that the VUs are served by the BSs with nearest distnce on either side of the highway, that is not blocked. For example, a vehicle on the top side of the highway can be served by a BS at the bottom side if either the BS at the bottom is nearer or the nearest BS at the top side is blocked.

Without loss of generality, we assume that the typical VU is located at the top side of the highway, i.e., $W_{t,l}<W_{b,l}$.  Here, we consider the beginning of the highway to be the point of origin and the VU lane as the x-axis, as shown in~Fig.~\ref{fig_lemma_2}. Let $\mathbf{b}_{t,1}, \mathbf{b}_{t,2},\dots$ and $\mathbf{b}_{b,1}, \mathbf{b}_{b,2},\dots$  be the sequences of x-axis coordinates of the BSs at the top and the bottom of the highway, respectively. Note that each coordinate is a point within a PPP, whose density is denoted by  $\lambda_{t,L}$ and $\lambda_{b,L}$, for the BS at the top and bottom, respectively. 


\begin{figure}[t]
    \centering
    \includegraphics*[width=0.8\columnwidth]{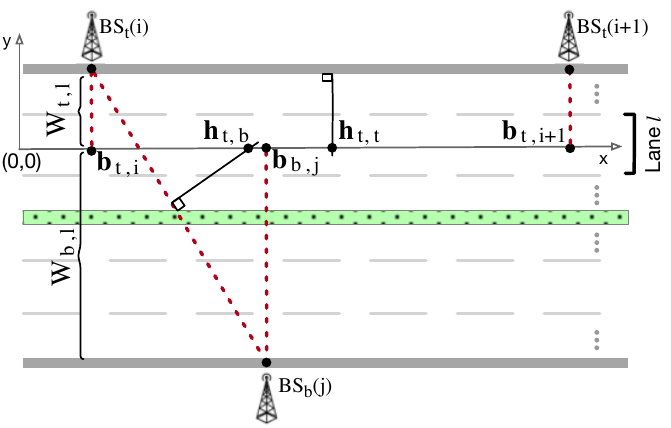}
    \caption{An example scenario illustrating a top to bottom side handover.}
    \vspace{-5mm}
    \label{fig_lemma_2}
\end{figure}

\begin{figure*}[!ht]
\begin{equation}
\setlength{\abovedisplayskip}{2pt}
\setlength{\belowdisplayskip}{2pt}
\label{eq_E_tb}
\begin{split}
 & \mathbb{E}\{\mathrm{N}_{tb}\}=\mathbb{E}\{\mathrm{N}_{bt}\}=\sum\limits_{k=0}^{\frac{N_{c}}{4}-1}\exp{\{-\lambda_{_{t,b,L}}(W_{b,l} a_k+\sqrt{W_{b,l}^2 a_k^2+W_{b,l}^2-W_{t,l}^2}\}}+
\\ & 
+\begin{cases}
\frac{N_{c}}{4}-\sum\limits_{k=0}^{\frac{N_{c}}{4}-1} 2 \exp{\{-\lambda_{_{t,b,L}}W_{t,l} a_{k}}\}  \sinh(\lambda_{_{t,b,L}} \sqrt{W_{t,l}^2 a_{k}^2+W_{t,l}^2-W_{b,l}^2})  & \sqrt{W_{b,l}^2-W_{t,l}^2}< W_{t,l} a_{0}
\\
\frac{N_{c}}{4}-\sum\limits_{k=M}^{\frac{N_{c}}{4}-1} 2 \exp{\{-\lambda_{_{t,b,L}}W_{t,l} a_{k}}\}  \sinh(\lambda_{_{t,b,L}} \sqrt{W_{t,l}^2 a_{k}^2+W_{t,l}^2-W_{b,l}^2}) & W_{t,l} a_{M-1} <\sqrt{W_{b,l}^2-W_{t,l}^2}< W_{t,l} a_{M}  \\ & \forall M=0,\dots, \frac{N_{c}}{4}-1
\\
\\
\frac{N_{c}}{4} & \sqrt{W_{b,l}^2-W_{t,l}^2}> W_{t,l} a_{_{\frac{N_{c}}{4}-1}}  
\end{cases}
\end{split}\tag{8}
\end{equation}
\hrulefill
\end{figure*}

For tractability, we first look into a section of the highway which is boxed between two BSs at the top side (i.e., between ${\rm BS}_t(i)$ and ${\rm BS}_t(i+1)$). Corrsponding to this box, there are a number of BSs on the bottom side of the highway, which are distributed according to a PPP.
Consequently, we first derive the probability that a typical VU handovers from a top BS to a bottom one and vice versa (see Lemmas~\ref{lem_P_tb} and \ref{lem_P_bt}). Then, considering one BS at the bottom of the highway,  we calculate the average number of beam switching and handover events from top to bottom and vice versa in Lemma~\ref{lem_E_tb}. Next, in Theorem~\ref{the_beam_switch}, we extend to the case that we have a number of PPP distributed BSs at the bottom side  while the VU traverses along the considered box. Finally, we generalize the results to the entire highway in Corollary~\ref{cor_beam_switching}.

\begin{lemma}
\label{lem_P_tb} 
Assuming that $\mathbf{b}_{b,j}$ refers to the first BS in the bottom side of the highway between $\mathbf{b}_{t,i}$ and $\mathbf{b}_{t,i+1}$, the probability of a handover to ${\rm BS}_b(j)$ while the VU traverses from ${\rm BS}_t(i)$ to ${\rm BS}_t(i+1)$ can be computed as
\begin{equation}
\label{eq_P_tb}
P_{tb}=\int_{0}^{\infty} (\lambda_{{t,L}}+\lambda_{{b,L}}) e^{-(\lambda_{{t,L}}+\lambda_{{b,L}})(x+\frac{W_{b,l}^2-W_{t,l}^2}{x})} dx.
\end{equation}
\begin{IEEEproof}
Define $\mathbf{h}_{t,t}$ as the handover point between ${\rm BS}_t(i)$ and ${\rm BS}_t(i+1)$. As mentioned in Section~\ref{ss:single_side}, the handover occurs when the distance between the VU and the serving BS is equal to the distance between the VU and the next LoS BS. Hence,
 \begin{equation}\label{eq_h_tt}
 \mathbf{h}_{t,t}=(\mathbf{b}_{t,i}+\mathbf{b}_{t,i+1})/2.
 \end{equation}
 
Let us denote $\mathbf{h}_{t,b}$ as the point of the handover between ${\rm BS}_t(i)$ and ${\rm BS}_b(j)$. Since $ (\mathbf{h}_{t,b}-\mathbf{b}_{t,i})^2+W_{t,l}^2=(\mathbf{b}_{b,j}-\mathbf{h}_{t,b})^2+W_{b,l}^2$, we can calculate $\mathbf{h}_{t,b}$ as

 \begin{equation} \label{eq_h_tb}
 \mathbf{h}_{t,b}=\dfrac{W_{b,l}^2-W_{t,l}^2+(\mathbf{b}_{b,j}-\mathbf{b}_{t,i})(\mathbf{b}_{b,j}+\mathbf{b}_{t,i})}{2(\mathbf{b}_{b,j}-\mathbf{b}_{t,i})}
	 \end{equation} 

Furthermore, the probability of a handover to ${\rm BS}_b(j)$ while the VU traverses from ${\rm BS}_t(i)$ to ${\rm BS}_t(i+1)$ is equivalent to the probability that $\mathbf{h}_{t,b}$  is before  $\mathbf{h}_{t,t}$, i.e., $\mathbf{h}_{t,b}<\mathbf{h}_{t,t}$ (see Fig.~\ref{fig_lemma_2}). So considering  \eqref{eq_h_tt} and \eqref{eq_h_tb},

 \begin{align*}
 P_{tb}\!=\!\mathrm{P}\big\{ \mathbf{h}_{t,b}\!<\!\mathbf{h}_{t,t}\big\} \!=\! \mathrm{P} \Big\{(\mathbf{b}_{t,i+1}\!-\!\mathbf{b}_{b,j})(\mathbf{b}_{b,j}-\mathbf{b}_{t,i})>
  W_{b,l}^2-W_{t,l}^2)\Big\}
 \end{align*}
Let us define $\mathbf{x}=\mathbf{b}_{t,i+1}-\mathbf{b}_{b,j}$ and $\mathbf{y}=\mathbf{b}_{b,j}-\mathbf{b}_{t,i}$. Then $\mathbf{x}$ and $\mathbf{y}$ are both i.i.d. random variables which follow an exponential distribution with density $\lambda_{t,L}+\lambda_{b,L}$. So  $P_{tb}$ is expressed as:

 \begin{align*}
  P_{tb}=&\mathrm{P} \Big\{\ \mathbf{x}\mathbf{y} >W_{b,l}^2-W_{t,l}^2 \Big\}=\int_{0}^{\infty}\int_{\frac{W_{b,l}^2-W_{t,l}^2}{x} }^{\infty} f_{\mathbf{x}\mathbf{y}}(x,y) dy dx \\ =& \int_{0}^{\infty} (\lambda_{t,L}+\lambda_{b,L}) e^{-(\lambda_{t,L}+\lambda_{b,L})(x+\frac{W_{b,l}^2-W_{t,l}^2}{x})} dx   
 \end{align*}
\end{IEEEproof}
\end{lemma}

\begin{lemma}
\label{lem_P_bt}
Assume that $\mathbf{b}_{b,j+1}$ falls between $\mathbf{b}_{b,j}$ and $\mathbf{b}_{t,i+1}$.  Then, the probability of a handover to ${\rm BS}_t(i)$ while traversing from the ${\rm BS}_b(j)$ to ${\rm BS}_b(j+1)$ is equal to
\begin{equation}
\label{eq_P_bt}
P_{bt}=1-\int_{0}^{\infty} (\lambda_{t,L}+\lambda_{b,L}) e^{-(\lambda_{t,L}+\lambda_{b,L})(x+\frac{W_{b,l}^2-W_{t,l}^2}{x})} dx,
\end{equation}
\begin{IEEEproof}
The proof is omitted due to the similarity to the proof of Lemma~\ref{lem_P_tb}.
\end{IEEEproof}
\end{lemma}

\begin{lemma}
\label{lem_E_tb}
The average number of beam switching events for a typical VU which handovers from a top side BS to a bottom side BS (i.e., the interval of $\mathbf{b}_{t,i} $ to $\mathbf{b}_{b,j}$ in Fig.\ref{fig_lemma_2})  is given by Eq.~\eqref{eq_E_tb}, where $a_k=\tan(\frac{\pi}{N_{c}}+\frac{2k\pi}{N_{c}}) $, $\lambda_{_{t,b,L}}=\lambda_{t,L}+\lambda_{b,L}$. The same expression can be used if the VU handovers from a bottom side BS to a top one.

 \begin{IEEEproof}
See Appendix \ref{App_lem_E_tb}.
 \end{IEEEproof}
\end{lemma}

\begin{theorem}
\label{the_beam_switch}
Considering a section of the highway enclosed  between two neighboring BSs at the top of the highway with distance $\mathbf{d}_{t}$, the average number of beam switching events $ \mathrm{NS}_{t,\mathrm{box}}$ and the average number of handovers events $ \mathrm{NH}_{t,\mathrm{box}}$ while the typical VU moves along the road restricted in this box are respectively equal to

\begin{figure*}[ht!] 
    \begin{subfigure}{.245\textwidth}
        \includegraphics*[width=1\columnwidth]{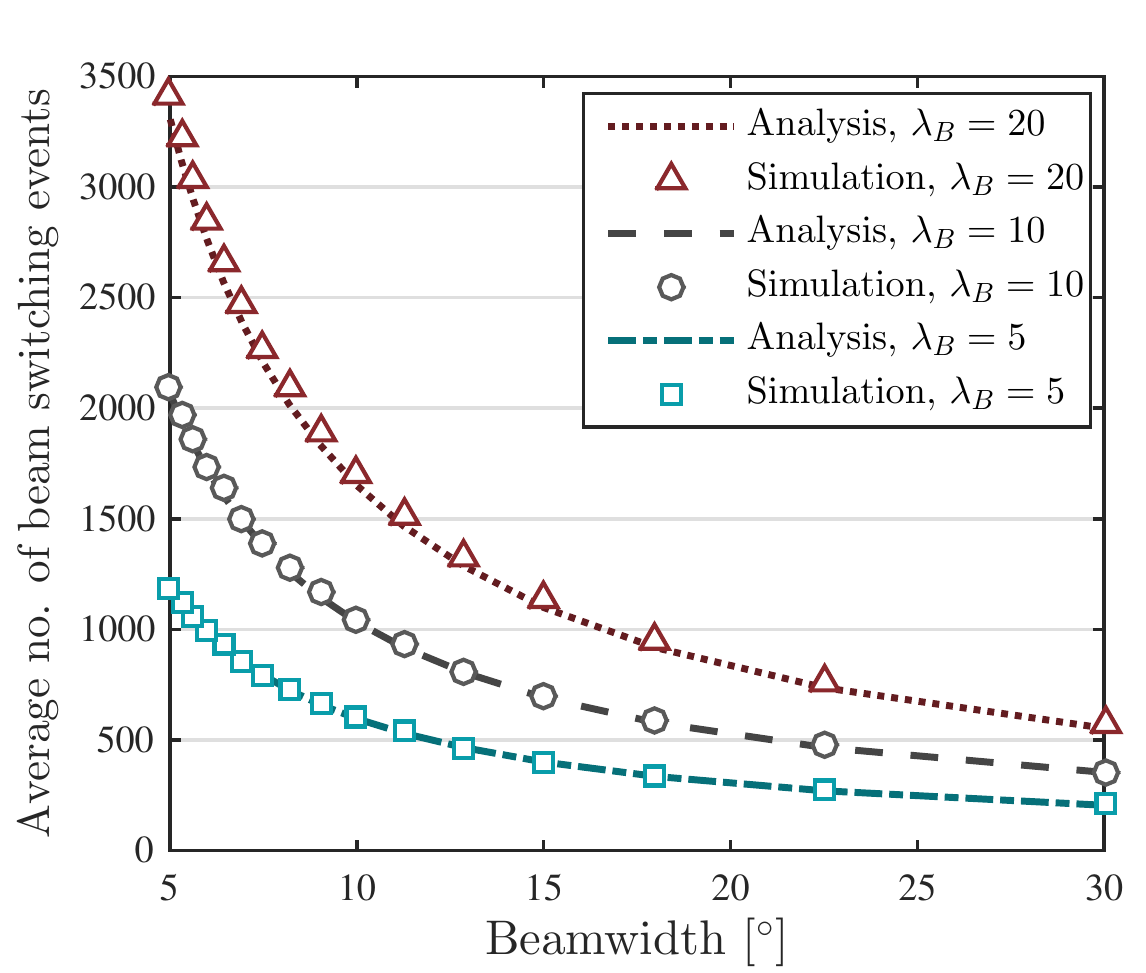}
        \caption{ }
        \label{beamswitching_vs_beam}
    \end{subfigure}
    \begin{subfigure}{.245\textwidth}
        \includegraphics*[width=1\columnwidth]{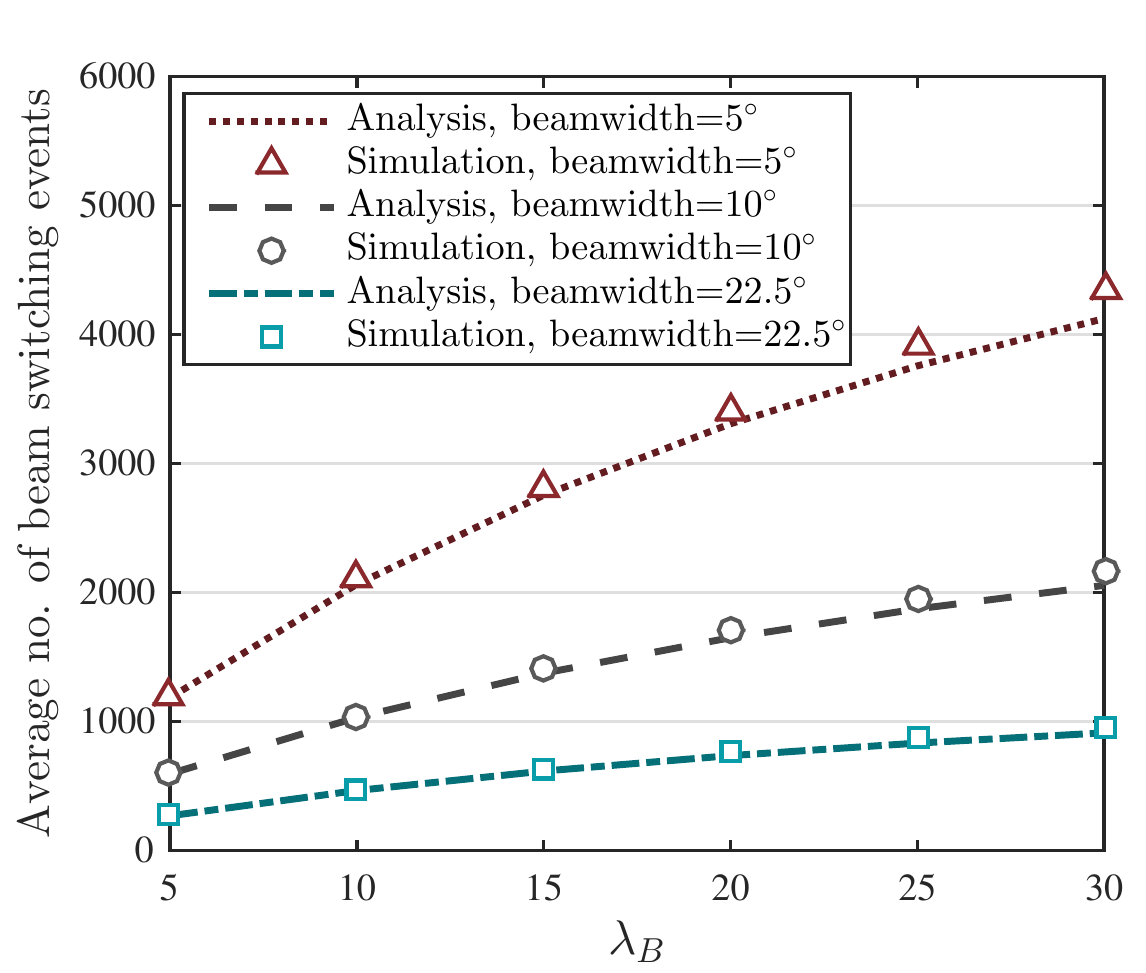}
        \caption{}
        \label{beamswitching_vs_lambda}
    \end{subfigure}
    \begin{subfigure}{.245\textwidth}
        \includegraphics*[width=1\columnwidth]{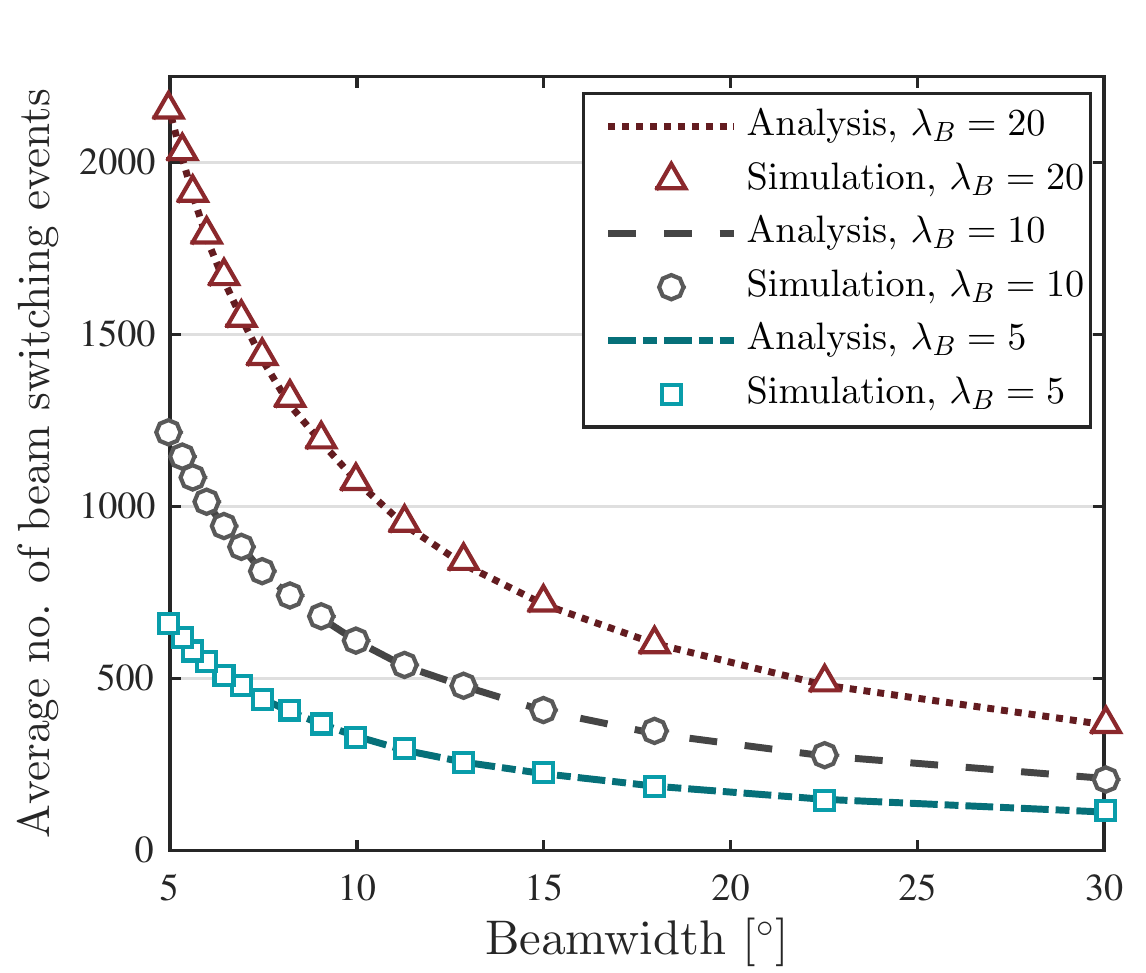}
        \caption{ }
        \label{beamswitching_vs_beam_one_side}
    \end{subfigure}
    \begin{subfigure}{.245\textwidth}
        \includegraphics*[width=1\columnwidth]{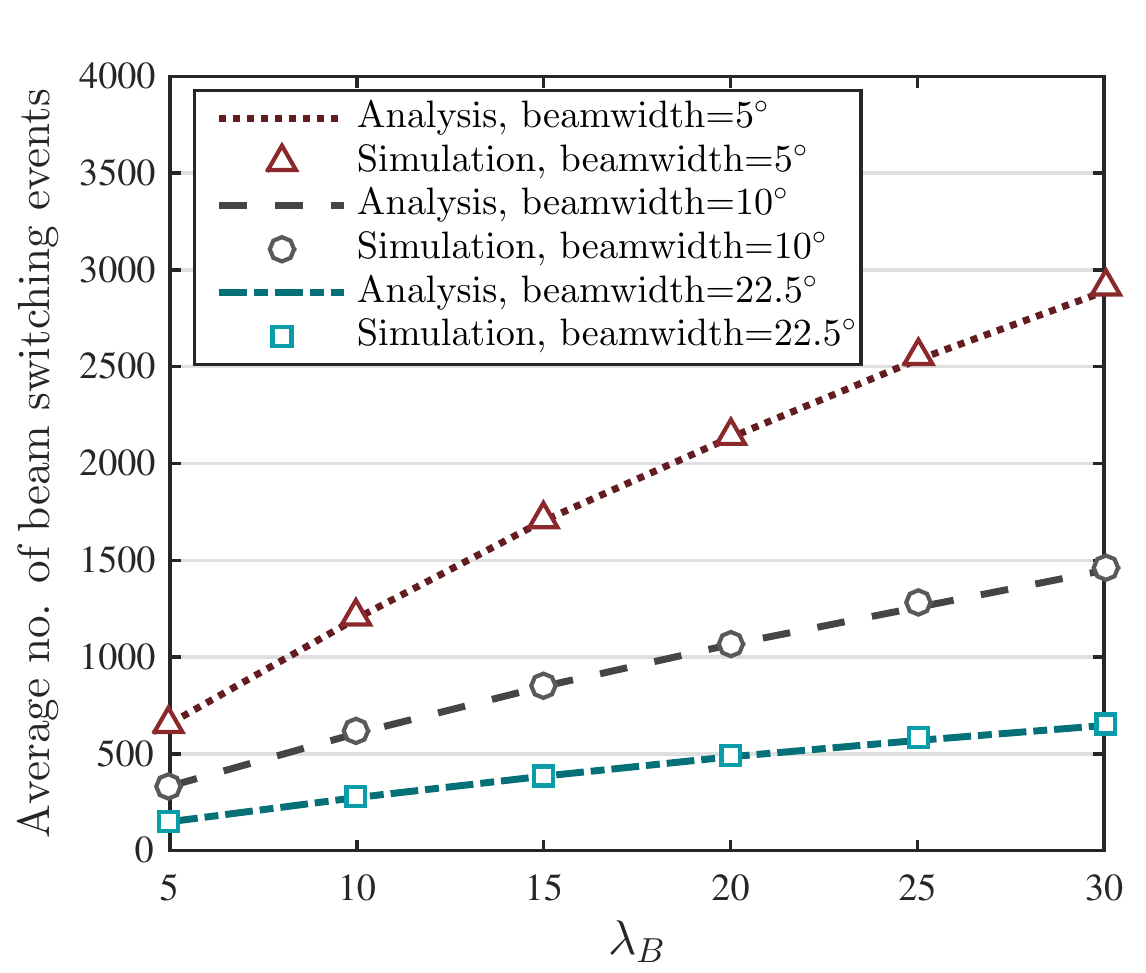}
        \caption{}
        \label{beamswitching_vs_lambda_one_side}
    \end{subfigure}
     \caption{The analysis and simulation of the average number of beam switching events versus the beamwidth  and the density of the BSs for double-side deployment, i.e., (a) and (b), as well as single-side deployment, i.e., (c) and (d).}
\label{beamswitching}
\end{figure*}
%
\begin{align}
\setcounter{equation}{8}
& \mathbb{E}\{\mathrm{NS}_{t,\mathrm{box}}\}=\frac{\lambda_{t,L}}{\lambda_{_{t,b,L}}}\sum\limits_{k=0}^{\frac{N_{c}}{4}-1} e^{-2W_{t,l} \lambda_{_{t,b,L}} a_{_k}}\notag\\&
+P_{tt}\sum\limits_{m=1}^{2}\sum\limits_{k=0}^{\frac{N_{c}}{4}-1} \dfrac{ \lambda_{t,L}\lambda_{b,L}(2W_{t,l}a_{_k})^{2-m} }{(2-m)!(\lambda_{_{t,b,L}})^m}e^{-2W_{t,l} \lambda_{_{t,b,L}} a_{_k}}\notag\\&
+\dfrac{\lambda_{t,L}\lambda_{b,L}}{(\lambda_{_{t,b,L}})^{2}}P_{tb}(\mathbb{E}\{\mathrm{N}_{tb}\}+\mathbb{E}\{\mathrm{N}_{bt}\})\notag \\&
+\sum\limits_{n_{_b}=2}^{\infty}\sum\limits_{m=1}^{n_{_b}+1}\sum\limits_{k=0}^{\frac{N_{c}}{4}-1} \dfrac{(P_{tt})^{n_{_b}}\lambda_{t,L}(\lambda_{b,L})^{n_{_b}}(2W_{t,l}a_{_k})^{n_{_b}-m+1} }{(n_{_b}-m+1)!(\lambda_{_{t,b,L}})^m}e^{-2W_{t,l} \lambda_{_{t,b,L}} a_{_k}}\notag\\&
+\sum\limits_{n_{_b}=2}^{\infty}\dfrac{\lambda_{t,L}(\lambda_{b,L})^{n_{_b}}}{(\lambda_{_{t,b,L}})^{n_{_b}+1}}\Bigg[ \Big(\mathbb{E}\{\mathrm{N}_{tb}\} + \mathbb{E}\{\mathrm{N}_{bt}\}\Big)\notag\\&\Big(\sum\limits_{n=1}^{n_{_b}-1} P_{tt}^{n-1}P_{tb}P_{bt}+P_{tt}^{n_{_b}-1}P_{tb}\Big)\notag
\\ & +\sum\limits_{n_{_s}=2}^{n_{_b}-1}\Big(\mathbb{E}\{\mathrm{N}_{tb}\}+
   (n_{_s}-1)\mathbb{E}\{\mathrm{N}_{b}(\mathbf{d}_b)\} + \mathbb{E}\{\mathrm{N}_{bt}\}\Big)\notag\\& \Big( \sum\limits_{n=1}^{n_{_b}-n_{_s}}P_{tt}^{n-1} P_{tb} P_{bb}^{n_{_s}-1}P_{bt}+ P_{tt}^{n_{_b}-n_{_s}} P_{tb} P_{bb}^{n_{_s}-1}\Big) +\notag\\& \Big(\mathbb{E}\{\mathrm{N}_{tb}\} + (n_{_b}-1)\mathbb{E}\{\mathrm{N}_{b}(\mathbf{d}_b)\}+ \mathbb{E}\{\mathrm{N}_{bt}\}\Big)P_{tb}P_{bb}^{n_{_b}-1} \Bigg] \label{eq1_th_beam_switch}
\end{align}
and  
\begin{align}
& \mathbb{E}\{\mathrm{NH}_{t,\mathrm{box}}\}=\frac{\lambda_{t,L}}{\lambda_{_{t,b,L}}}+P_{tt}\dfrac{\lambda_{t,L}\lambda_{b,L}}{(\lambda_{_{t,b,L}})^{2}}\notag\\&+2P_{tb}\dfrac{\lambda_{t,L}\lambda_{b,L}}{(\lambda_{_{t,b,L}})^{2}}+\sum\limits_{n_{_b}=2}^{\infty}(P_{tt})^{n_{_b}}\dfrac{\lambda_{t,L}(\lambda_{b,L})^{n_{_b}}}{(\lambda_{_{t,b,L}})^{n_{_b}+1}}\notag\\&+\sum\limits_{n_{_b}=2}^{\infty}\dfrac{\lambda_{t,L}(\lambda_{b,L})^{n_{_b}}}{(\lambda_{_{t,b,L}})^{n_{_b}+1}}\Bigg[ 2\Big(\sum\limits_{n=1}^{n_{_b}-1} P_{tt}^{n-1}P_{tb}P_{bt}+P_{tt}^{n_{_b}-1}P_{tb}\Big)\notag
\\ & +\sum\limits_{n_{_s}=2}^{n_{_b}-1}\Big(1+n_{_s}\Big) \Big( \sum\limits_{n=1}^{n_{_b}-n_{_s}}P_{tt}^{n-1} P_{tb} P_{bb}^{n_{_s}-1}P_{bt}+ P_{tt}^{n_{_b}-n_{_s}} P_{tb} P_{bb}^{n_{_s}-1}\Big) +\notag\\& \Big(n_{_b}+1\Big)P_{tb}P_{bb}^{n_{_b}-1} \Bigg] \label{eq2_th_beam_switch}
\end{align}
where $a_k=\tan(\frac{\pi}{N_{c}}+\frac{2k\pi}{N_{c}}) $, $\lambda_{_{t,b,L}}=\lambda_{t,L}+\lambda_{b,L}$, $P_{tb}$ is as derived in lemma \ref{lem_P_tb}, $P_{bt}$ is given in lemma~\ref{lem_P_bt}, $P_{tt}=1-P_{tb}$, and $P_{bb}=1-P_{bt}$. Moreover, $\mathbb{E}\{\mathrm{N}_{tb}\}$ and $\mathbb{E}\{\mathrm{N}_{bt}\}$ are given in lemma $\ref{lem_E_tb}$ and  $\mathbb{E}\{\mathrm{N}_{b}(\mathbf{d}_b)\}$ is given in Theorem~\ref{the:beam_switch1}.
\begin{IEEEproof}
See Appendix \ref{App_the_beam_switch}.
\end{IEEEproof}
\end{theorem}

\begin{corollary}
\label{cor_beam_switching}
The average number of beam switching events and the average number of handover events  for the typical VU driving on the $l^{\rm th}$ lane at the top side of the highway could be expressed as
 $$\mathrm{BSN}=\lambda_{t,L} L_h\mathbb{E}\{\mathrm{NS}_{t,\mathrm{box}}\}$$ and
$$\mathrm{HON}=(\lambda_{t,L} L_h-1)\mathbb{E}\{\mathrm{NH}_{t,\mathrm{box}}\},$$
where $\lambda_{t,L}$ is the density of LoS BSs at the top side of the highway, $L_h$ is the length of the highway, and $\mathbb{E}\{\mathrm{NS}_{t,\mathrm{box}}\}$ and $\mathbb{E}\{\mathrm{NS}_{t,\mathrm{box}}\}$ are given in Theorem\ref{the_beam_switch}.

\begin{IEEEproof}
Let $ \mathbf{n}_{_{B,t}}$ be the number of LOS BSs at the top side of the highway, which is distributed according to a Poisson distribution with density $\lambda_{t,L}L_h$, then thenumber of beam switching and handover events along the highway, respectively would be 


\begin{align*}
& {\rm BSN} = \mathbb{E}\{ \mathbf{n}_{_{B,t}}\}\mathbb{E}\{\mathrm{NS}_{t,\mathrm{box}}\} = \mathbb{E}\{ \mathbf{n}_{_{B,t}}\}\mathbb{E}\{\mathrm{NS}_{t,\mathrm{box}}\}= \\ & L_h \lambda_{t,L}\mathbb{E}\{\mathrm{NS}_{t,\mathrm{box}}\}.
\end{align*}
and
%
%
%
\begin{align*}
{\rm HON} =(\mathbb{E}\{ \mathbf{n}_{_{B,t}}\}-1)\mathbb{E}\{\mathrm{NH}_{t,\mathrm{box}}\}= (L_h \lambda_{t,L}-1)\mathbb{E}\{\mathrm{NH}_{t,\mathrm{box}}\}
\end{align*}
\end{IEEEproof}
\end{corollary}

\section{Evaluation}

 In this section, we evaluate the performance of mmWave vehicular communication under high mobility. In particular, we study the system performance in terms of  number of beam switching events, the signaling overhead, and beam sojourn time. Furthermore, we verify the accuracy of our analyses by providing both numerical evaluation and Monte-Carlo simulation for the every scenario under study. In our evaluation, we consider a $ 4 $-lane divided highway, in which the right lanes on each direction are dedicated to large vehicles, as shown in Fig.~\ref{fig:sys_model}. Table \ref{tab:param} provides the default values for the simulations and numerical evaluations. 
 




\begin{table}[b]
\scriptsize
\caption{Simulation and analysis parameters }
\vspace{-2mm}
\centering
\begin{tabular}{|l|c|}
\hline
{\bf Parameter}								& {\bf Value} 					\\ \hline\hline
System Bandwidth							& $ 1 $GHz				\\ \hline
Transmit power								& $27$dBm				\\ \hline
Thermal noise					& $ -174\  \text{dBm/Hz} $			\\ \hline
Carrier Frequency ($f$)                                                               & $ 28 $GHz             \\ \hline
LS Pathloss Component ($C$) \cite{ }                                         & $ 20 \log_{10} (4\pi f / c)$dB		\\ \hline
Antenna	 height 						                                                     & $10$m				\\ \hline
The width of each highway lane 					                               & $ 3.7$m				\\ \hline
The length of the highway  ($L_h$)					                            & $10$km				\\ \hline
The number of the highway lanes  ($N_L$)                                 & $4$				\\ \hline
The number of blockage highway lanes                                      & $2$				\\ \hline
The number of  VN highway lanes                                               & $2$				\\ \hline
The density of the VNs in each lane ($\lambda_{v,l} \ \forall l$)  & $1$ per km				\\ \hline
The density of the blockages in each lane                                   & $0.1$ per km				\\ \hline
The length of each blockage($\tau_{0}$)	                                    & $9$m		\\ \hline
The SSB duration ($\tau_{ss}$)	                                    & $5$ms		\\ \hline
The symbol duration ($\tau_{sym}$)	                                    & $0.125$ms		\\ \hline
\end{tabular}
\label{tab:param}
\vspace{-4mm}
\end{table}


\begin{figure*}[!ht] 
\centering
    \begin{subfigure}{.3\textwidth}
        \includegraphics*[width=1\columnwidth]{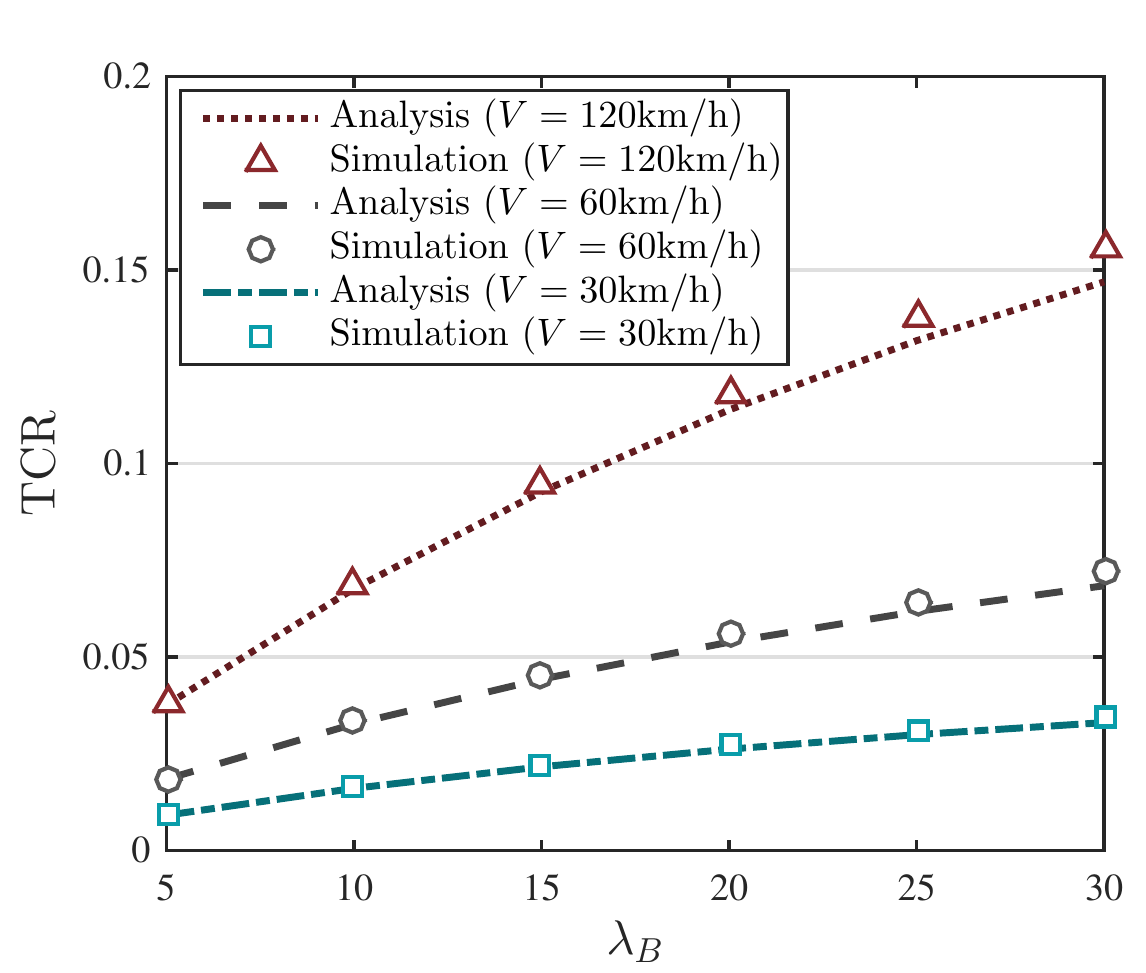}
        \caption{}
        \label{TCR_lambda_72}        
    \end{subfigure}
    \begin{subfigure}{.3\textwidth}
        \includegraphics*[width=1\columnwidth]{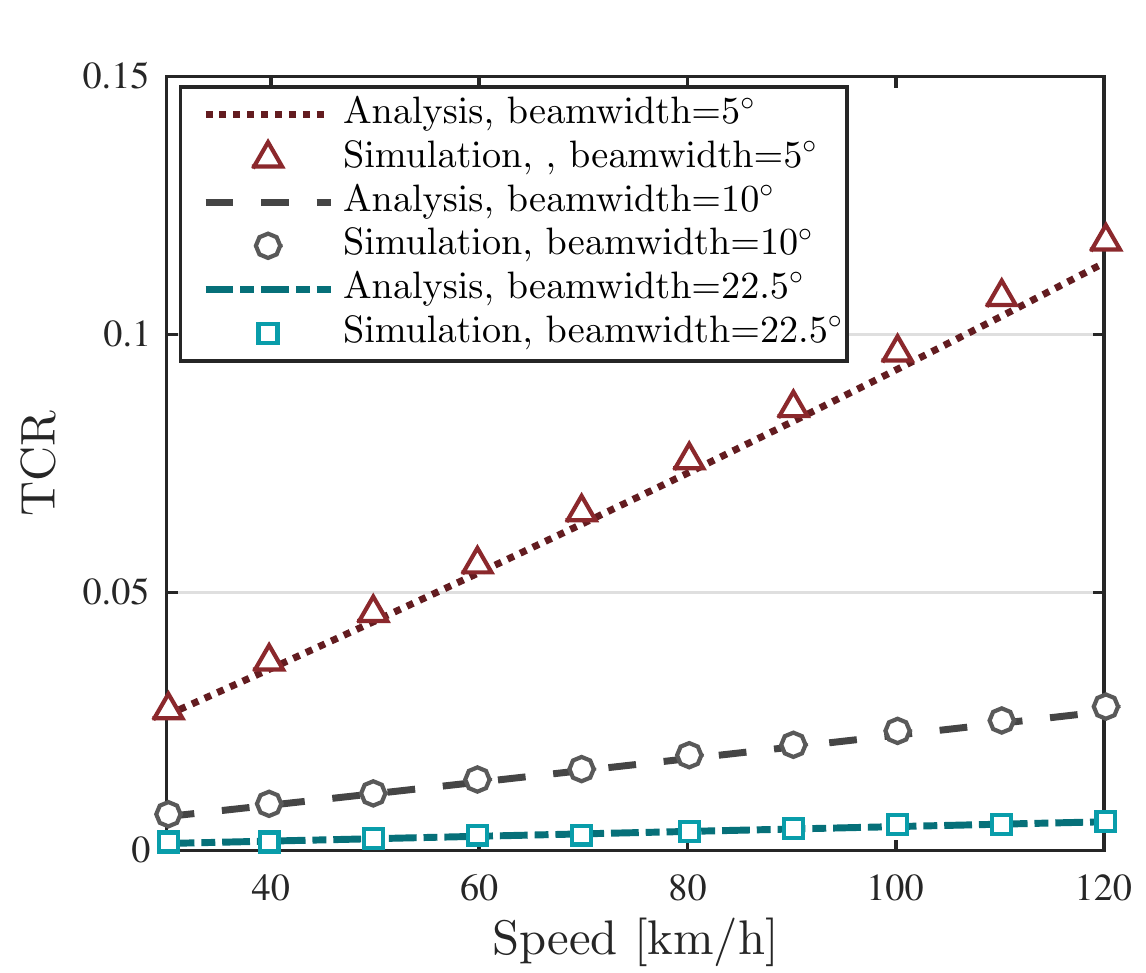}
        \caption{}
        \label{TCR_speed_20}        
    \end{subfigure}
    \begin{subfigure}{.3\textwidth}
        \includegraphics*[width=1\columnwidth]{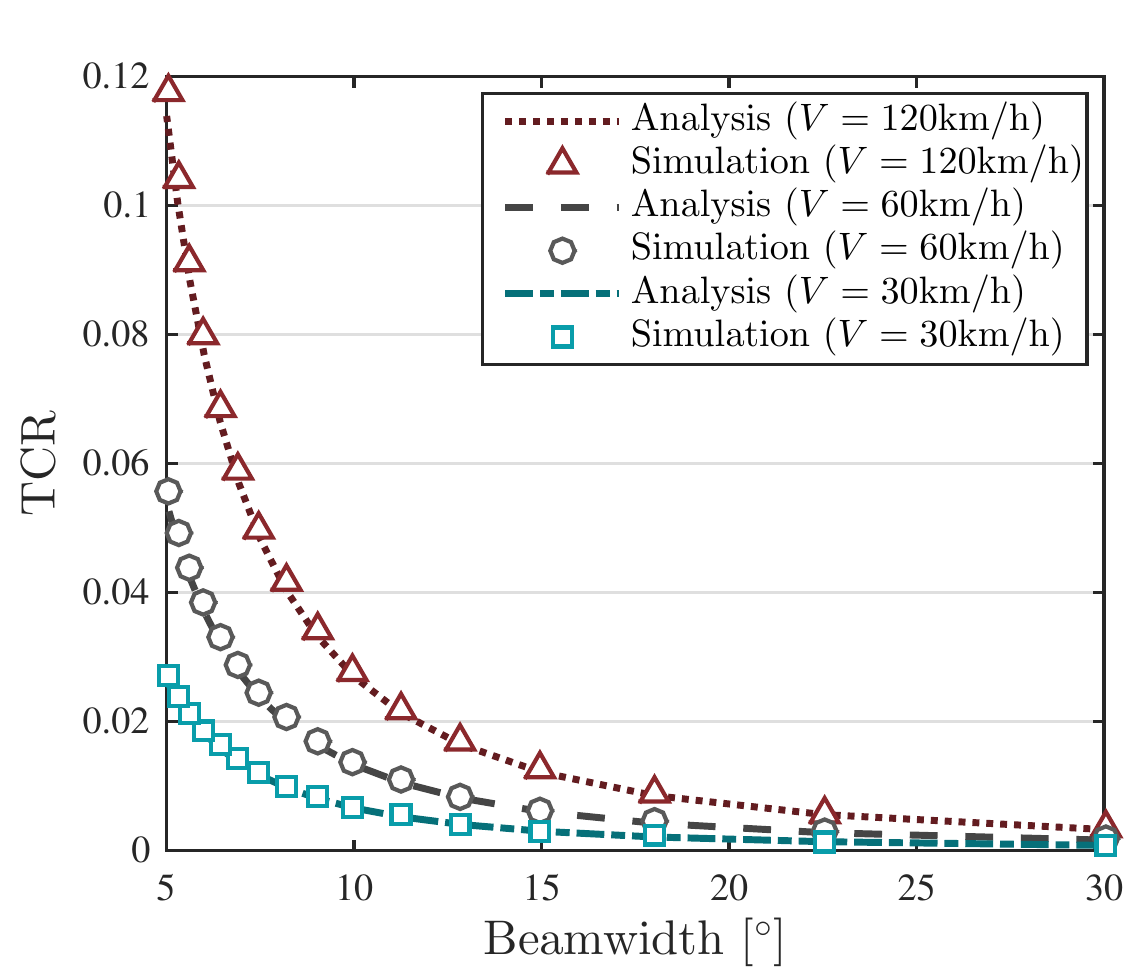}
        \caption{}
        \label{TCR_beamwidth_20}                
    \end{subfigure}
    \caption{The analysis and simulation of the TCR versus: (a) the density of the BSs, (b) the speed of the vehicles, and (c) the beamwidth.}
    \label{TCR}
\end{figure*}

\subsection{beam switching analysis}
\label{ss:beamswitching}
In this section, we investigate impact of beamwidth and BS density on the average number of the beam switching events.

In Fig.~\ref{beamswitching}, we show how the number of beam switching events is impacted by the beamwidth (see Fig.~\ref{beamswitching_vs_beam}) and the density of BSs (see Fig.~\ref{beamswitching_vs_lambda}) on a double-side deployment along the highway. Fig.~\ref{beamswitching_vs_beam} demonstrates that the number of the beam switching events is inversely related to the beamwidth at the BSs. This is expected since the coverage area of a beam is effectively reduced as the beam get narrower. Fig.~\ref{beamswitching_vs_lambda} shows that the beam switching events become more frequent in denser networks. The densification's impact is twofold: $ (i) $ the number of handover events increases with the number of BSs, and $ (ii) $ the number of beam switching events also increases since the densifications leads to selection of the beams that cover smaller areas of the road which also provide lower pathloss. We will elaborate on this further in Section~\ref{ss:TCR}. Note the slight deviation between simulation and analysis in Fig.~\ref{beamswitching_vs_lambda}. The deviation stems from the approximation of the Erlang distribution by an exponential one in Section~\ref{ss:divided}, where we extended the derivations of the single-side deployment to the double-side deployment. This is confirmed by looking at the results in Figs.~\ref{beamswitching_vs_lambda_one_side}~and~\ref{beamswitching_vs_beam_one_side}, where we observe that the analysis and simulations are completely matching.

\subsection{Beam training overhead}
\label{ss:TCR}
Beam management can induce significant overhead to mmWave networks because the time spent in beam alignment effectively reduces the available time for data communication. To this aim, in this subsection, we introduce the training to connectivity ratio (TCR) metric which is in essence an indicator of the portion of the time spent on beam alignment in reference to the total connectivity time, that is the time allocated to actual data communication. The following is the formal definition of TCR, based on the description provided in Section~\ref{s:background}:

\begin{equation}
    \mathrm{TCR}=\dfrac{T_{\mathrm{ho}}+ T_{\mathrm{bswitch}}}{L_h V- (T_{\mathrm{ho}}+ T_{\mathrm{bswitch}})},
\end{equation}
where $V$ is the speed of the VU, and  $T_{\mathrm{ho}}$ and $T_{\mathrm{bswitch}}$ are the total time spent for beam sweeping after each handover between BSs and after moving from the coverage area of one beam to the other, respectively. Both values can be computed as follows~\cite{giordani2018tutorial}:
\begin{align}
    T_{ho} [ms] = |ho| \times (\frac{|codebook_{bs}| \times |codebook_{vu}|} {64} \times \tau_{ss} ), \label{T_handover}
\end{align}
where $|ho|$ is the number of handovers, and $|codebook_{vu}|$ and $|codebook_{vu}|$ are the codebook size of the BS and VU, respectively. Finally, $ \tau_{ss}$ is the duration of synchronization signal burst as defined in Section~\ref{s:background}.

\begin{align}
T_{bswitch} [ms] =  & |bswitch| \times \\ \nonumber
& (|codebook_{bs}| \times |codebook_{vu}| \times  \tau_{sym} ) \label{T_Beam}
\end{align}
where $|bswitch|$ is the number of handover events and $\tau_{sym}$ is the duration of an OFDM symbol. Having defined TCR, we now elaborate on the impact of BS density, vehicular speed, and beamwidth in Figs.~\ref{TCR}.

 Fig.~\ref{TCR_lambda_72} shows the value of TCR versus the density of the BSs under $5^\circ$ beamwidth. Here we observe two general trends: $ (i) $ TCR increases with BS density, and $(ii)$ TCR grows with speed of the VU. As the density of BS grows, the number of the handover and beam switching points also grows, but it is not intuitively apparent which one contributes more to the overall overhead. Our evaluation results demonstrate that on average in all cases $\sim 90\%$ of the total overhead is due to beam switching. Of course, as mentioned in Section~\ref{ss:beamswitching}. the number of handovers increases with the BS density. However, the densification also leads to the use of beams which are closer to the BS and provide smaller coverage area but higher SNR. This is better observed in Fig.~\ref{fig:beamswitching}. We can see in the figure that the beam intersecting with the x-axis at $W_{i,l} a_{_{0}}$ provides smaller coverage area than $W_{i,l} a_{_{1}}$, that is, the vehicles are mostly served by beams which are (almost) perpendicular to the highway as opposed to those that cover the corners. The speed, on the other hand, reduces the time spent in coverage of each BS and every beam. However, the time required for beam alignment does not decrease with the speed. Hence, we see higher TCR values at higher vehicular speeds, see Fig.~\ref{TCR_speed_20}. Last, we observe how beamwidth affects the TCR in Fig.~\ref{TCR_beamwidth_20}. It can be seen that the TCR reduction with beamwidth is non-linear, which is similar to the behavior observed for BS density and speed. It should be noted that overhead of beam management can easily reach to $10\%$ that is a large burden on the operators networks if not carefully planned.

\begin{figure*}[!ht] 
\centering
    \begin{subfigure}{.3\textwidth}
        \includegraphics*[width=1\columnwidth]{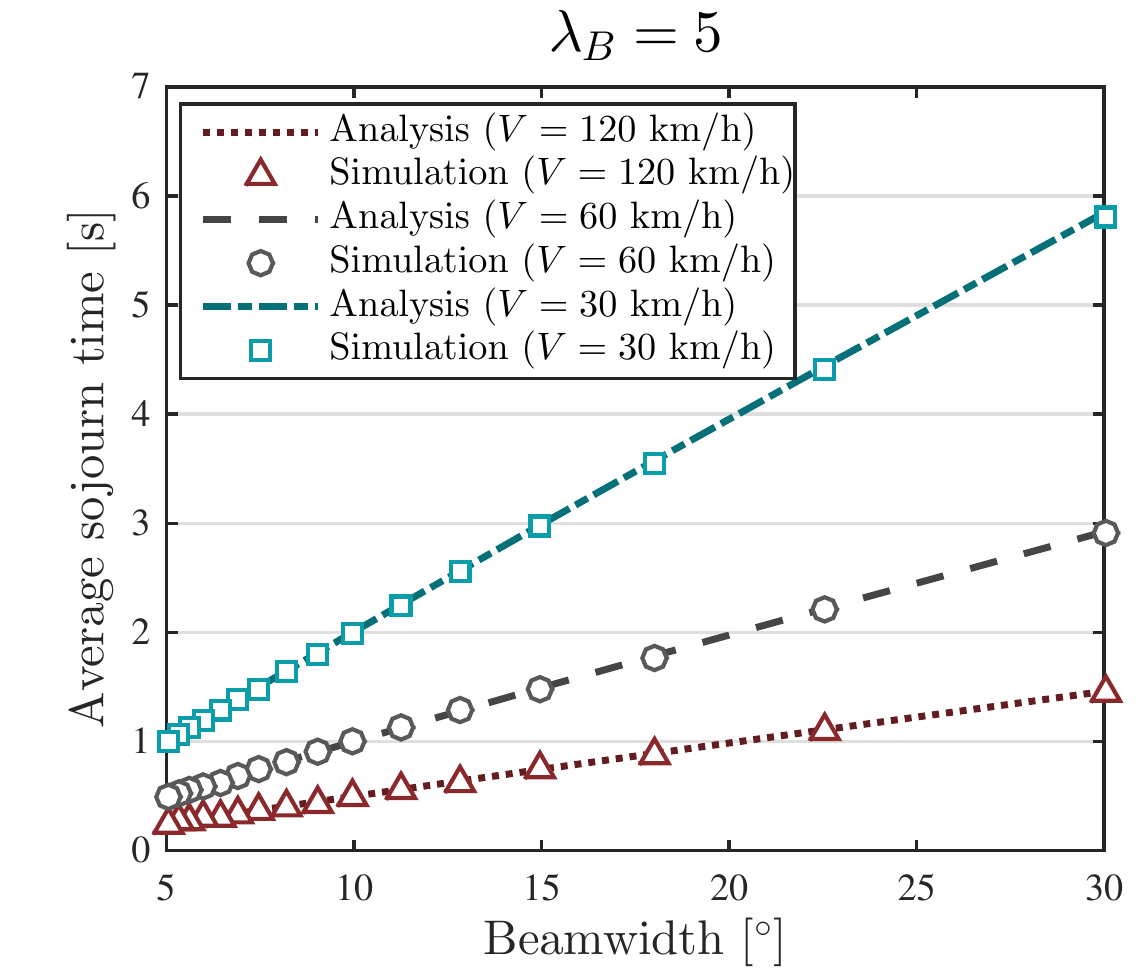}
        \caption{}\label{sojoun_time_lambda_5}
    \end{subfigure}
    \begin{subfigure}{.3\textwidth}
        \includegraphics*[width=1\columnwidth]{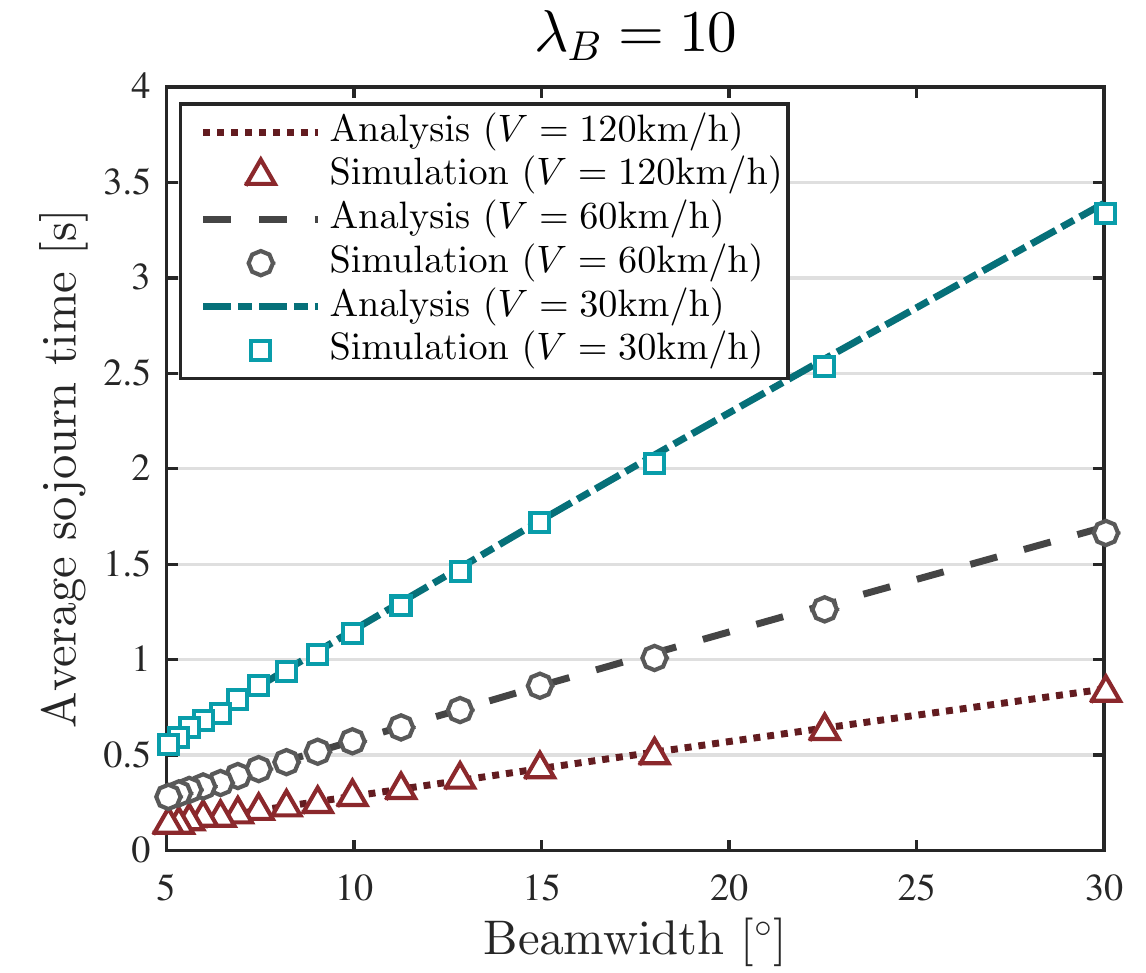}
        \caption{}\label{sojoun_time_lambda_10}
    \end{subfigure}
    \begin{subfigure}{.3\textwidth}
        \includegraphics*[width=1\columnwidth]{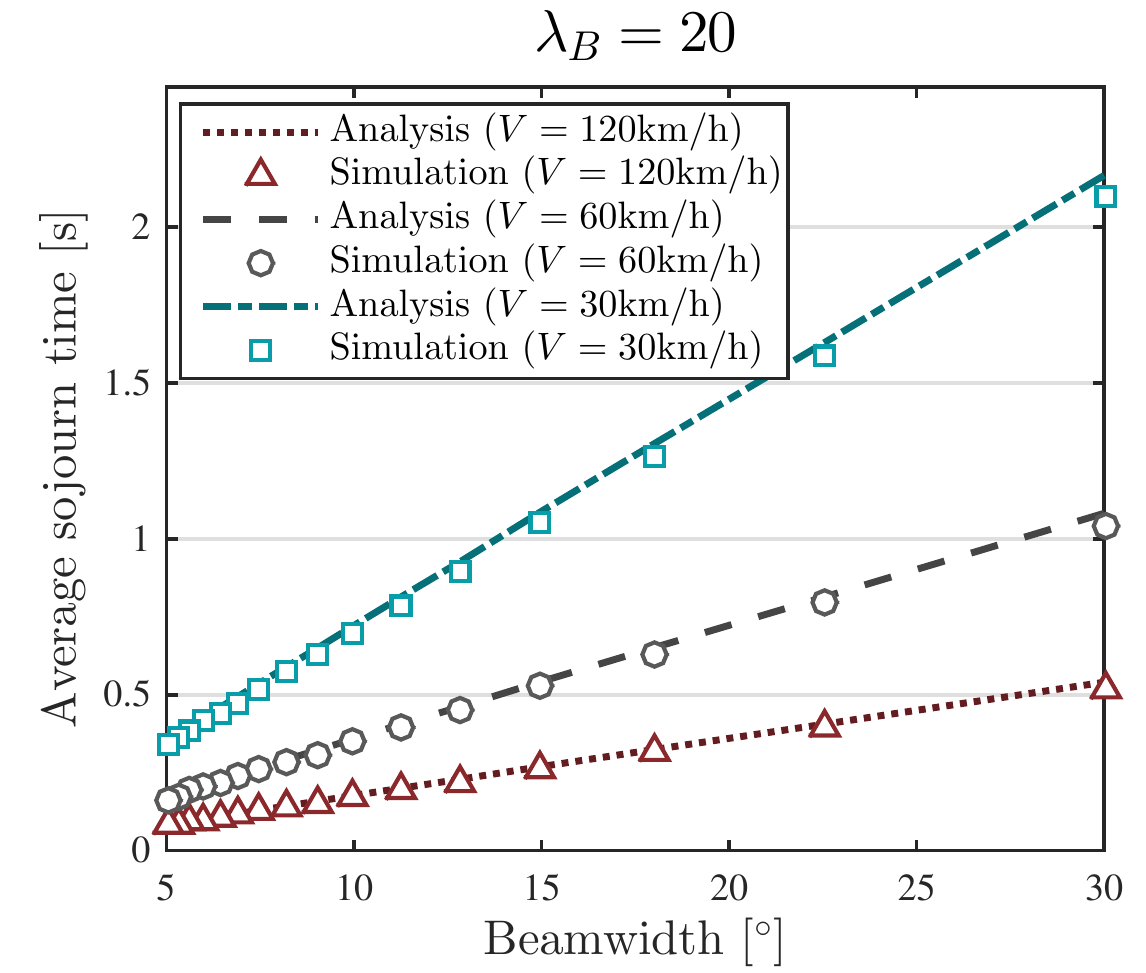}
        \caption{}\label{sojoun_time_lambda_20}
    \end{subfigure}
    \caption{The analysis and simulation of  the  average sojourn time versus beamwidth of the BSs for speeds  $30, 60, 120 \  \mathrm{km/h}$ (a): $\lambda_B=5$. (b): $\lambda_B=10$, (c): $\lambda_B=20$   }
\label{sojourn_time}
\end{figure*}

\subsection{Design insights}
As mentioned, the 3GPP specification leaves selection of many parameters to the operators. In this section, we demonstrate how our analysis can be used to make suitable design choices. 

In Fig.~\ref{sojourn_time}, we illustrate the average sojourn time of the VU in each beam versus the beamwidth of the BSs. We define sojourn time as the time a vehicle stays within the coverage of a beam. Recall that the periodicity of the beam sweeping ($T_{ss}$) and CSI-RS reports ($T_{csi}$) is left to the network operators to decide. Knowing the average sojourn time, the operators can derive suitable values for these parameters. As a rule of thumb, the frequency of beam alignment should be less than the beam sojourn time, otherwise, the vehicles may transition from one beam to the other before the system can detect the misalignment, that eventually renders the mmWave connection very intermittent. 

Focusing on the results of beam sojourn evaluation, we can see that: $(i)$ the beam sojourn time reduces as the BS density increases, $ (ii) $ the beam sojourn time reduces as the speed grows, and $ (iii) $ the sojourn time grows with the beamwidth. These figures emphasize on the required network agility for high mobility scenarios. For example, the average sojourn time of $78$ ms demands for very frequent beam sweeping and CSI-RS reports to ensure that the mmWave connectivity remains uninterrupted. 

It should be also noted that the TCR results in Fig.~\ref{TCR} are helpful to network operators for link budget estimation. Knowing the TCR, the operators can estimate the effective capacity of the network for a given deployment density with relation to the typical mobility speed of the vehicles in that area. Note that the overhead of beam management can vary up to $80\%$ in different vehicular speeds (see Fig.~\ref{TCR_speed_20}), which significantly affects the network planning. 


\section{Conclusion}
\label{s:conclusion}

In this article, we have provided a stochastic geometry-based framework to analyze beam management in mmWave vehicular networks. In particular, we have derived several closed-form expressions for the number of beam switching and handover events. Via extensive evaluation, we have provided a thorough picture of major impacting factor on beam management. We have studied the overall system overhead as well as the beam sojourn time. Using this results, we provided design insights which can be leveraged for selection of suitable standard defined parameters. 

\bibliographystyle{IEEEtran}


\bibliography{biblio}

\begin{thebibliography}{10}
\providecommand{\url}[1]{#1}
\csname url@samestyle\endcsname
\providecommand{\newblock}{\relax}
\providecommand{\bibinfo}[2]{#2}
\providecommand{\BIBentrySTDinterwordspacing}{\spaceskip=0pt\relax}
\providecommand{\BIBentryALTinterwordstretchfactor}{4}
\providecommand{\BIBentryALTinterwordspacing}{\spaceskip=\fontdimen2\font plus
\BIBentryALTinterwordstretchfactor\fontdimen3\font minus
  \fontdimen4\font\relax}
\providecommand{\BIBforeignlanguage}[2]{{%
\expandafter\ifx\csname l@#1\endcsname\relax
\typeout{** WARNING: IEEEtran.bst: No hyphenation pattern has been}%
\typeout{** loaded for the language `#1'. Using the pattern for}%
\typeout{** the default language instead.}%
\else
\language=\csname l@#1\endcsname
\fi
#2}}
\providecommand{\BIBdecl}{\relax}
\BIBdecl

\bibitem{Bao:2015vl}
W.~Bao and B.~Liang, ``{Stochastic Geometric Analysis of User Mobility in
  Heterogeneous Wireless Networks},'' \emph{IEEE Journal on Selected Areas in
  Communications}, vol.~33, pp. 2212--2225, 2015.

\bibitem{ElSawy:2017bn}
H.~ElSawy, A.~K. Sultan-Salem, M.-S. Alouini, and M.~Z. Win, ``{Modeling and
  Analysis of Cellular Networks Using Stochastic Geometry - A Tutorial.}''
  \emph{IEEE Communications Surveys and Tutorials}, vol.~19, no.~1, pp.
  167--203, 2017.

\bibitem{zeinalpour2018effect}
Z.~Zeinalpour-Yazdi, S.~Jalali, and H.~V. Poor, ``Effect of backhaul
  constraints on uplink femtocell networks,'' \emph{IEEE Transactions on
  Vehicular Technology}, vol.~67, no.~10, pp. 9931--9946, 2018.

\bibitem{cho2018v2x}
Y.~J. Cho, K.~Huang, and C.-B. Chae, ``V2x downlink coverage analysis with a
  realistic urban vehicular model,'' in \emph{2018 IEEE Globecom Workshops (GC
  Wkshps)}.\hskip 1em plus 0.5em minus 0.4em\relax IEEE, 2018, pp. 1--6.

\bibitem{chetlur2019coverage}
V.~V. Chetlur and H.~S. Dhillon, ``Coverage and rate analysis of downlink
  cellular vehicle-to-everything (c-v2x) communication,'' \emph{arXiv preprint
  arXiv:1901.09236}, 2019.

\bibitem{sial2019stochastic}
M.~N. Sial, Y.~Deng, J.~Ahmed, A.~Nallanathan, and M.~Dohler, ``Stochastic
  geometry modeling of cellular v2x communication over shared channels,''
  \emph{IEEE Transactions on Vehicular Technology}, 2019.

\bibitem{wang2018mmwave}
Y.~Wang, K.~Venugopal, A.~F. Molisch, and R.~W. Heath, ``Mmwave
  vehicle-to-infrastructure communication: Analysis of urban microcellular
  networks,'' \emph{IEEE Transactions on Vehicular Technology}, vol.~67, no.~8,
  pp. 7086--7100, 2018.

\bibitem{yi2019modeling}
W.~Yi, Y.~Liu, Y.~Deng, A.~Nallanathan, and R.~W. Heath, ``Modeling and
  analysis of mmwave v2x networks with vehicular platoon systems,'' \emph{IEEE
  Journal on Selected Areas in Communications}, 2019.

\bibitem{tassi2017modeling}
A.~Tassi, M.~Egan, R.~J. Piechocki, and A.~Nix, ``Modeling and design of
  millimeter-wave networks for highway vehicular communication,'' \emph{IEEE
  Transactions on Vehicular Technology}, vol.~66, no.~12, pp. 10\,676--10\,691,
  2017.

\bibitem{giordani2018coverage}
M.~Giordani, M.~Rebato, A.~Zanella, and M.~Zorzi, ``Coverage and connectivity
  analysis of millimeter wave vehicular networks,'' \emph{Ad Hoc Networks},
  vol.~80, pp. 158--171, 2018.

\bibitem{giordani2018tutorial}
M.~Giordani, M.~Polese, A.~Roy, D.~Castor, and M.~Zorzi, ``A tutorial on beam
  management for 3gpp nr at mmwave frequencies,'' \emph{IEEE Communications
  Surveys \& Tutorials}, vol.~21, no.~1, pp. 173--196, 2018.

\bibitem{3GPP38.802}
{3GPP}, ``Study on new radio (nr) access technology - physical layer aspects,''
  TR 38.802, 2018, v15.0.0.

\bibitem{3GPP38.213}
------, ``{NR - Physical layer procedures for control},'' TS 38.213, 2018,
  v15.0.0.

\bibitem{3GPP38.331}
------, ``{NR - Radio Resource Control (RRC) protocol specification},'' TS
  38.331, 2018, v15.0.0.

\bibitem{elsawy2013stochastic}
H.~ElSawy, E.~Hossain, and M.~Haenggi, ``Stochastic geometry for modeling,
  analysis, and design of multi-tier and cognitive cellular wireless networks:
  A survey,'' \emph{IEEE Communications Surveys \& Tutorials}, vol.~15, no.~3,
  pp. 996--1019, 2013.

\bibitem{baccelli2010stochastic}
F.~Baccelli, B.~B{\l}aszczyszyn \emph{et~al.}, ``Stochastic geometry and
  wireless networks: Volume ii applications,'' \emph{Foundations and
  Trends{\textregistered} in Networking}, vol.~4, no. 1--2, pp. 1--312, 2010.

\bibitem{andrews2016modeling}
J.~G. Andrews, T.~Bai, M.~N. Kulkarni, A.~Alkhateeb, A.~K. Gupta, and R.~W.
  Heath, ``Modeling and analyzing millimeter wave cellular systems,''
  \emph{IEEE Transactions on Communications}, vol.~65, no.~1, pp. 403--430,
  2016.

\end{thebibliography}

\appendices
\section{The proof of  Lemma \ref{lem_E_tb}}
\label{App_lem_E_tb}
The amount of  $\mathrm{N}_{tb}$ is equal to the number of beam switching events from ${\rm BS}_t(i)$ to the handover point ($\mathrm{N}_{th}$) plus the number of beam switching events from the handover point to ${\rm BS}_b(j)$ ($\mathrm{N}_{hb}$). Thus, we split this problem to two parts where in the first part the average of $\mathrm{N}_{th}$  is computed and then in the second part we derive the average of $\mathrm{N}_{hb}$.
We define $\mathbf{h}_{t,b}$ as the point of the handover between ${\rm BS}_t(i)$ and ${\rm BS}_b(j)$. 
 \begin{figure}[b]
 \centering
\includegraphics*[width=.9\columnwidth]{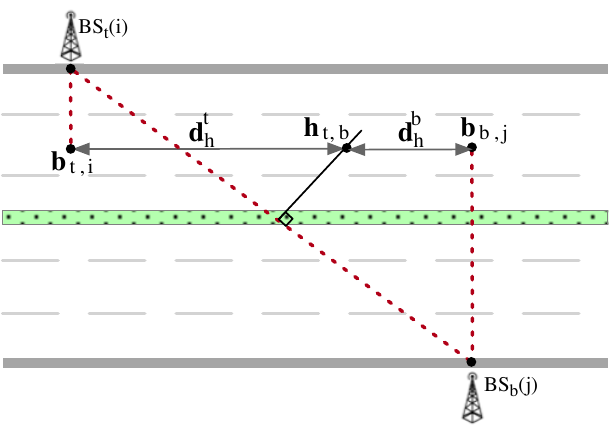}
\caption{An example scenario illustrating the loctaion of top to bottom side handover point.}
 \label{fig_lemma4}
\end{figure}
 Moreover, we define $\mathbf{d}_{h}^{t}=\mathbf{h}_{t,b}-\mathbf{b}_{t,i}$ and $\mathbf{d}_{h}^{b}=\mathbf{b}_{b,j}-\mathbf{h}_{t,b}$ (Fig.\ref{fig_lemma4}), thus similar to the proof of lemma \ref{lem:num_4} one can proof that,
 \begin{equation}
 \label{Num_th}
 \mathrm{N}_{th}=\begin{cases}
0  & 0<\mathbf{d}_{h}^{t}<  W_{t,l} a_{_{0}} 
\\
1 & W_{t,l} a_{_{0}}  <\mathbf{d}_{h}^{t}< W_{t,l} a_{_{1}}
\\
\vdots
\\
k & W_{t,l} a_{_{k-1}}  <\mathbf{d}_{h}^{t}< W_{t,l}a_{_{k}}
\\
\vdots 
\\
\frac{N_c}{4} & W_{t,l} a_{_{\frac{N_c}{4}-1} } <\mathbf{d}_{h}^{t}
\end{cases}
 \end{equation}
 and
  \begin{equation}
 \mathrm{N}_{hb}=\begin{cases}
0  & 0<\mathbf{d}_{h}^{b}<  W_{b,l} a_{_{0}} 
\\
1 & W_{b,l} a_{_{0}}  <\mathbf{d}_{h}^{b}< W_{b,l} a_{_{1}}
\\
\vdots
\\
k & W_{b,l} a_{_{k-1}}  <\mathbf{d}_{h}^{b}< W_{b,l} a_{_{k}}
\\
\vdots 
\\
\frac{N_c}{4} & W_{b,l} a_{_{\frac{N_c}{4}-1} } <\mathbf{d}_{h}^{b}
\end{cases}
 \end{equation}
 where $a_k=\tan(\frac{\pi}{N_c}+\frac{2k\pi}{N_c})$.\\ \\
1) Evaluation of $\mathbb{E}\{\mathrm{N}_{th}\}$:
 Using \eqref{eq_h_tb}, we have
 \begin{equation}
 \label{eq_d_ht}
 \mathbf{d}_{h}^{t}=\dfrac{W_{b,l}^2-W_{t,l}^2}{2(\mathbf{b}_{b,j}-\mathbf{b}_{t,i})}+\dfrac{\mathbf{b}_{b,j}-\mathbf{b}_{t,i}}{2}.
 \end{equation}
 We define $\mathbf{x}=\mathbf{b}_{b,j}-\mathbf{b}_{t,i}$, thus  $\mathbf{x}\sim \exp (\lambda_{_{t,b,L}})$, where $\lambda_{_{t,b,L}}=\lambda_{t,L}+\lambda_{b,L}$,  therefore

 \begin{equation}
 \label{f_d_th}
 \begin{split}
 & f_{_{\mathbf{d}_{h}^{t}}}(y)=\lambda_{_{t,b,L}} (\dfrac{y}{\sqrt{y^2+W_{t,l}^2-W_{b,l}^2}}+1)\\ &e^{ -\lambda_{_{t,b,L}}(y+\sqrt{y^2+W_{t,l}^2-W_{b,l}^2})} u(y-\sqrt{W_{b,l}^2-W_{t,l}^2}) \\ &+ \lambda_{_{t,b,L}} (\dfrac{y}{\sqrt{y^2+W_{t,l}^2-W_{b,l}^2}}-1)\\ &e^{ -\lambda_{_{t,b,L}}(y-\sqrt{y^2+W_{t,l}^2-W_{b,l}^2})}  u(y-\sqrt{W_{b,l}^2-W_{t,l}^2})
 \end{split}
 \end{equation}
 where $\lambda_{_{t,b,L}}=\lambda_{t,L}+\lambda_{b,L}$.
 So
 \begin{align} 
 & \mathbb{E}\{\mathrm{N}_{th}\}=\int \mathrm{N}_{th}(y)f_{_{\mathbf{d}_{h}^{t}}}(y) dy= \notag \\ &\int_{\sqrt{W_{b,l}^2-W_{t,l}^2}}^{\infty} \mathrm{N}_{th}(y) \Big( \lambda_{_{t,b,L}} (\dfrac{y}{\sqrt{y^2+W_{t,l}^2-W_{b,l}^2}}+1)\\ &e^{ -\lambda_{_{t,b,L}}B(y+\sqrt{y^2+W_{t,l}^2-W_{b,l}^2})} \notag\\ &+ \lambda_{_{t,b,L}} (\dfrac{y}{\sqrt{y^2+W_{t,l}^2-W_{b,l}^2}}-1)\\ &e^{ -\lambda_{_{t,b,L}}(y-\sqrt{y^2+W_{t,l}^2-W_{b,l}^2})} \Big) dy. \notag
 \end{align}
 Considering \eqref{Num_th}, we would have multi states for 
 $\mathbb{E}\{\mathrm{N}_{th}\}$, i.e.,  $\mathbb{E}\{\mathrm{N}_{th}\}$
 is a piecewise function of $W_{t,l}$ and $W_{b,l}$. 
 
 $\bullet$ If $\sqrt{W_{b,l}^2-W_{t,l}^2}< W_{t,l} a_{_0}$:
 \begin{align*} 
 & \mathbb{E}\{\mathrm{N}_{th}\}= \sum\limits_{k=1}^{\frac{N_c}{4}-1} k \int_{W_{t,l} a_{_{k-1}}}^{W_{t,l} a_{_{k}}}\\ &\Big(  \lambda_{_{t,b,L}} (\dfrac{y}{\sqrt{y^2+W_{t,l}^2-W_{b,l}^2}}+1)\\ &e^{ -\lambda_{_{t,b,L}}(y+\sqrt{y^2+W_{t,l}^2-W_{b,l}^2})} \\ &+ \lambda_{_{t,b,L}} (\dfrac{y}{\sqrt{y^2+W_{t,l}^2-W_{b,l}^2}}-1)\\ &e^{ -\lambda_{_{t,b,L}}(y-\sqrt{y^2+W_{t,l}^2-W_{b,l}^2})} \Big) dy \\ &+\frac{N_c}{4}\int_{W_{t,l} a_{_{\frac{N_c}{4}-1}}}^{\infty}\Big( \lambda_{_{t,b,L}} (\dfrac{y}{\sqrt{y^2+W_{t,l}^2-W_{b,l}^2}}+1)\\ &e^{ -\lambda_{_{t,b,L}}(y+\sqrt{y^2+W_{t,l}^2-W_{b,l}^2})} \\ &+ \lambda_{_{t,b,L}} (\dfrac{y}{\sqrt{y^2+W_{t,l}^2-W_{b,l}^2}}-1)\\ &e^{ -\lambda_{_{t,b,L}}(y-\sqrt{y^2+W_{t,l}^2-W_{b,l}^2})}\Big) dy \\ &=\sum\limits_{k=1}^{\frac{N_c}{4}-1} k \Big[e^{ -\lambda_{_{t,b,L}}(y+\sqrt{y^2+W_{t,l}^2-W_{b,l}^2})}\Big]_{W_{t,l} a_{_k}}^{W_{t,l} a_{_{k-1}}}+\\ &+\frac{N_c}{4}\Big[e^{ -\lambda_{_{t,b,L}}(y+\sqrt{y^2+W_{t,l}^2-W_{b,l}^2})}\Big]_{\infty}^{W_{t,l} a_{_{\frac{N_c}{4}-1}}}+\\ &+
 \sum\limits_{k=1}^{\frac{N_c}{4}-1} k \Big[e^{ -\lambda_{_{t,b,L}}(y-\sqrt{y^2+W_{t,l}^2-W_{b,l}^2})}\Big]_{W_{t,l} a_{_{k-1}}}^{W_{t,l} a_{_{k}}}+\\ &+\frac{N_c}{4}\Big[e^{ -\lambda_{_{t,b,L}}(y-\sqrt{y^2+W_{t,l}^2-W_{b,l}^2})}\Big]_{W_{t,l} a_{_{\frac{N_c}{4}-1}}}^{\infty}\\ & = \frac{N_c}{4}-2\sum\limits_{k=0}^{\frac{N_c}{4}-1} \Big( e^{-\lambda_{_{t,b,L}}W_{t,l} a_{k}} \\ & \sinh(\lambda_{_{t,b,L}} \sqrt{W_{t,l}^2 a_{k}^2+W_{t,l}^2-W_{b,l}^2}) \Big)
\end{align*}  
where the last equality is resulted by rearranging the statements.
\\
$\bullet$ If $W_{t,l} a_{_{M-1}}<\sqrt{W_{b,l}^2-W_{t,l}^2} <W_{t,l} a_{_{M}}$ for  one $M\in \{1,\dots,\frac{N_c}{4}-1\} $,   	
\begin{align*}
&\mathbb{E}\{\mathrm{N}_{th}\}=M \int_{\sqrt{W_{b,l}^2-W_{t,l}^2}}^{W_{t,l} a_{_{M}}} \\ &\Big(  \lambda_{_{t,b,L}} (\dfrac{y}{\sqrt{y^2+W_{t,l}^2-W_{b,l}^2}}+1)e^{ -\lambda_{_{t,b,L}}(y+\sqrt{y^2+W_{t,l}^2-W_{b,l}^2})} \\ &+ \lambda_{_{t,b,L}} (\dfrac{y}{\sqrt{y^2+W_{t,l}^2-W_{b,l}^2}}-1)e^{ -\lambda_{_{t,b,L}}(y-\sqrt{y^2+W_{t,l}^2-W_{b,l}^2})} \Big) dy+\\ &+\sum_{k=M+1}^{\frac{N_c}{4}-1} k \int_{W_{t,l} a_{_{k-1}}}^{W_{t,l} a_{_{k}}} \\ & \Big(  \lambda_{_{t,b,L}} (\dfrac{y}{\sqrt{y^2+W_{t,l}^2-W_{b,l}^2}}+1)e^{ -\lambda_{_{t,b,L}}(y+\sqrt{y^2+W_{t,l}^2-W_{b,l}^2})} \\ &+ \lambda_{_{t,b,L}} (\dfrac{y}{\sqrt{y^2+W_{t,l}^2-W_{b,l}^2}}-1)e^{ -\lambda_{_{t,b,L}}(y-\sqrt{y^2+W_{t,l}^2-W_{b,l}^2})} \Big) dy+\\ & \frac{N_c}{4}\int_{W_{t,l} a_{_{\frac{N_c}{4}-1}}}^{\infty}\\ &\Big(  \lambda_{_{t,b,L}} (\dfrac{y}{\sqrt{y^2+W_{t,l}^2-W_{b,l}^2}}+1)e^{ -\lambda_{_{t,b,L}}(y+\sqrt{y^2+W_{t,l}^2-W_{b,l}^2})} \\ &+ \lambda_{_{t,b,L}} (\dfrac{y}{\sqrt{y^2+W_{t,l}^2-W_{b,l}^2}}-1)e^{ -\lambda_{_{t,b,L}}(y-\sqrt{y^2+W_{t,l}^2-W_{b,l}^2})} \Big) dy\\ &=\frac{N_c}{4}-\sum\limits_{k=M}^{\frac{N_c}{4}-1} 2 e^{-\lambda_{_{t,b,L}}W_{t,l} a_{k}}  \sinh(\lambda_{_{t,b,L}} \sqrt{W_{t,l}^2 a_{k}^2+W_{t,l}^2-W_{b,l}^2})
\end{align*}
where the last equality is resulted by solving the integrals similar to the previous state.
\\
$\bullet$  Finally, if $\sqrt{W_{b,l}^2-W_{t,l}^2}>W_{t,l} a_{\frac{N_c}{4}-1}$:
\begin{align*}
 & \mathbb{E}\{\mathrm{N}_{th}\}=\int_{\sqrt{W_{b,l}^2-W_{t,l}^2}}^{\infty}\\ &
 \Big(  \lambda_{_{t,b,L}} (\dfrac{y}{\sqrt{y^2+W_{t,l}^2-W_{b,l}^2}}+1)e^{ -\lambda_{_{t,b,L}}(y+\sqrt{y^2+W_{t,l}^2-W_{b,l}^2})} \\ &+ \lambda_{_{t,b,L}} (\dfrac{y}{\sqrt{y^2+W_{t,l}^2-W_{b,l}^2}}-1)e^{ -\lambda_{_{t,b,L}}(y-\sqrt{y^2+W_{t,l}^2-W_{b,l}^2})} \Big) dy\\&=\frac{N_c}{4}\Big[e^{ -\lambda_{_{t,b,L}}(y+\sqrt{y^2+W_{t,l}^2-W_{b,l}^2})}\Big]_{\infty}^{\sqrt{W_{b,l}^2-W_{t,l}^2}}\\&+\frac{N_c}{4}\Big[e^{ -\lambda_{_{t,b,L}}(y-\sqrt{y^2+W_{t,l}^2-W_{b,l}^2})}\Big]_{\sqrt{W_{b,l}^2-W_{t,l}^2}}^{\infty}\\&=\frac{N_c}{4}
\end{align*}
 2) Evaluation of $\mathbb{E}\{\mathrm{N}_{hb}\}$:
We defined   $\mathbf{d}_{h}^{b}=\mathbf{b}_{b,j}-\mathbf{h}_{t,b}$, then using \eqref{eq_h_tb}, we have
 \begin{equation}
 \label{eq_d_hb}
 \mathbf{d}_{h}^{b}=-\dfrac{W_{b,l}^2-W_{t,l}^2}{2(\mathbf{b}_{b,j}-\mathbf{b}_{t,i})}+\dfrac{\mathbf{b}_{b,j}-\mathbf{b}_{t,i}}{2}
 \end{equation}
 Now defining  $\mathbf{x}=\mathbf{b}_{b,j}-\mathbf{b}_{t,i}$ (where $\mathbf{x}\sim \exp(\lambda_{_{t,b,L}})$ ) we have
 \begin{equation}
 \begin{split}
 & f_{_{\mathbf{d}_{h}^{b}}}(y)=\\&\lambda_{_{t,b,L}} (\dfrac{y}{\sqrt{y^2+W_{b,l}^2-W_{t,l}^2}}+1)e^{ -\lambda_{_{t,b,L}}(y+\sqrt{y^2+W_{b,l}^2-W_{t,l}^2})} u(y)
 \end{split}
 \end{equation}
Thus
\begin{align*}
&\mathbb{E}\{\mathrm{N}_{hb}\}=\sum\limits_{k=1}^{\frac{N_c}{4}-1} k \int_{W_{b,l}a_{_{k-1}}}^{W_{b,l}a_{_k}}\\& \lambda_{_{t,b,L}} (\dfrac{y}{\sqrt{y^2+W_{b,l}^2-W_{t,l}^2}}+1)e^{ -\lambda_{_{t,b,L}}(y+\sqrt{y^2+W_{b,l}^2-W_{t,l}^2})} dy \\&+\int_{W_{b,l}a_{_{\frac{N_c}{4}-1}}}^{\infty} \frac{N_c}{4}\lambda_{_{t,b,L}} (\dfrac{y}{\sqrt{y^2+W_{b,l}^2-W_{t,l}^2}}+1)\\&e^{ -\lambda_{_{t,b,L}}(y+\sqrt{y^2+W_{b,l}^2-W_{t,l}^2})} dy \\&=\sum\limits_{k=1}^{\frac{N_c}{4}-1} k \Big[e^{ -\lambda_{_{t,b,L}}(y+\sqrt{y^2+W_{b,l}^2-W_{t,l}^2})}\Big]_{W_{b,l}a_{_{k-1}}}^{W_{b,l}a_{_k}}\\&+\frac{N_c}{4}\Big[e^{ -\lambda_{_{t,b,L}}(y+\sqrt{y^2+W_{b,l}^2-W_{t,l}^2})}\Big]_{W_{b,l}a_{_{\frac{N_c}{4}-1}}}^{\infty}\\&=\sum\limits_{k=0}^{\frac{N_c}{4}-1}e^{-\lambda_{_{t,b,L}}(W_{b,l} a_k+\sqrt{W_{b,l}^2 a_k^2+W_{b,l}^2-W_{t,l}^2})}
\end{align*}
\\ 
 Considering that $\mathbb{E}\{\mathrm{N}_{\mathrm{tb}}\}=\mathbb{E}\{\mathrm{N}_{th}\}+\mathbb{E}\{\mathrm{N}_{hb}\}$,  \eqref{eq_E_tb} is obtained.
 \\
 Similarly, could be proved that $\mathbb{E}\{\mathrm{N}_{\mathrm{bt}}\}$ is also equal to \eqref{eq_E_tb}.
\section{The proof of theorem \ref{the_beam_switch}}
\label{App_the_beam_switch}
Assume that  $\mathbf{b}_{b,j},\dots, \mathbf{b}_{b,j+\mathbf{n}_{_{b}}-1}$ are  between $\mathbf{b}_{t,i}$
 and $\mathbf{b}_{t,i+1}$, and $\mathbf{n}_{_{b}} \sim \mathrm{poiss}(\lambda_{b,L} \mathbf{d}_{t})$, where $\mathbf{d}_{t}$ is the distance of  $\mathbf{b}_{t,i}$ to $\mathbf{b}_{t,i+1}$. In this appendix we calculate the  number of beam switching events ( $\mathrm{NS}_{t,\mathrm{box}}$) and the  number of handovers events, $ \mathrm{NH}_{t,\mathrm{box}}$, while the typical VU moves through the coverage area between  ${\rm BS}_t(i)$ and ${\rm BS}_t(i+1)$.
 
 We define $\mathbf{n}_{_{bv}}$ to be the number of the BSs on the bottom side of the highway which serve the VU during the movement from $\mathbf{b}_{t,i}$ to  $\mathbf{b}_{t,i+1}$, thus 
 $\mathbf{n}_{_{bv}} \in \{0,\dots,\mathbf{n}_{_{b}}\}$, and so
 \begin{align*}
  &\mathrm{NS}_{t,\mathrm{box}}(\mathbf{n}_{_{bv}},\mathbf{n}_{_{b}}, \mathbf{d}_{t})=
\\  &\begin{cases}
 \mathrm{N}_{t}(\mathbf{d}_t) & \mathbf{n}_{_{b}}\geq 0, \mathbf{n}_{_{bv}}=0 
 \\ \\
   (\mathbb{E}\{\mathrm{N}_{tb}\}+\mathbb{E}\{\mathrm{N}_{bt}\})  &
  \mathbf{n}_{_{b}}=1, \mathbf{n}_{_{bv}}=1
   \\ \\
(\mathbb{E}\{\mathrm{N}_{tb}\}+ 
 (\mathbf{n}_{_{bv}}-1)\mathbb{E}\{\mathrm{N}_{b}(\mathbf{d}_b)\} +\\ \mathbb{E}\{\mathrm{N}_{bt}\})
 &\mathbf{n}_{_{b}}\geq 2, \ 1\leq \mathbf{n}_{_{bv}} \leq \mathbf{n}_{_{b}}
  \end{cases}
\end{align*}  
\\ 
 Moreover, $\mathbf{n}_{_{b}}\sim \mathrm{poiss}(\lambda_{b,L}\mathbf{d}_{t})$ and 
$\mathbf{b}_{t,1}, \mathbf{b}_{t,2},\dots$  are the points of a PPP with density $\lambda_{t,L}$, thus $\mathbf{d}_{t} \sim \exp(\lambda_{t,L})$ and then the average of $ \mathrm{NS}_{t,\mathrm{box}}(\mathbf{n}_{_{bv}},\mathbf{n}_{_{b}}, \mathbf{d}_{t})$ is as follows
\begin{align}
&\mathbb{E}\{\mathrm{NS}_{t,\mathrm{box}}\} \notag \\ &=\mathbb{E}_{\mathbf{d}_{t}}{\{\mathbb{E}_{\mathbf{n}_{_{b}}} }\{ \mathbb{E}_{\mathbf{n}_{_{bv}}}\{\mathrm{NS}_{t,\mathrm{box}} | \mathbf{n}_{_{b}}=n_{b},\mathbf{d}_{t}=r  \}\}\} \notag\\ =&
\int_{0}^{\infty}\sum\limits_{n_{_b}=0}^{\infty}\mathbb{E}_{\mathbf{n}_{_{bv}}}\{\mathrm{NS}_{t,\mathrm{box}} | \mathbf{n}_{_{b}}=n_{_b},\mathbf{d}_{t}=r  \}\notag\\&\dfrac{e^{-\lambda_{b,L}r}(\lambda_{b,L}r)^{n_{_b}}}{{n_{_b}}!}\lambda_{t,L}e^{-\lambda_{t,L}r} dr
\label{eq_Num_tt}
 \end{align}
\\
On the other hand considering lemma \ref{lem_P_bt} and lemma  \ref{lem_P_tb}, the conditional probability function of $\mathbf{n}_{_{bv}}$ is as  Table \ref{tab_n_{_{bv}}} (for the deteails of derivations refer to Fig.~\ref{state_machine}),   where $P_{tb}$ and $P_{bt}$ are as defined in lemma \ref{lem_P_tb} and lemma \ref{lem_P_bt},
  $P_{tt}=1-P_{tb}$ and $P_{bb}=1-P_{bt}$. We should say that for simplicity, we approximate the distribution of the distance between two non-consecutive points by the Exponential distribution, rather than the Erlang distribution.
\begin{table*}
\scriptsize
\caption{The Conditional probability function of $\mathbf{n}_{_{bv}}$}
\centering
\begin{tabular}{c|cccc} 
 $\mathbf{n}_{s}$ & $0$ & $1$ & $2 \leq n_{_{bv}} \leq n_{_b}-1$ & $n_{_{bv}}=n_{_b}\geq 2$ \\ \hline
\\
$\mathrm{P}\{\mathbf{n}_{_{bv}}=n_{_{bv}}| \mathbf{n}_{_{b}}=0 \}$& $1$ & $0$ & $0$ & $0$\\ 
\\
$\mathrm{P}\{\mathbf{n}_{_{bv}}=n_{_{bv}}| \mathbf{n}_{_{b}}=1 \} $ & $P_{tt}$ &$P_{tb}$  &$0$ & $0$  \\ 
 \\
$\mathrm{P}\{\mathbf{n}_{_{bv}}=n_{_{bv}}| \mathbf{n}_{_{b}}=n_{_b} \geq 2\} $ &$P_{tt}^{n_{_b}}$   &$\sum\limits_{n=1}^{n_{_b}-1} P_{tt}^{n-1}P_{tb}P_{bt}+P_{tt}^{n_{_b}-1}P_{tb}$  & $\sum\limits_{n=1}^{n_{_b}-n_{_{bv}}} P_{tt}^{n-1} P_{tb} P_{bb}^{n_{_{bv}}-1}P_{bt}+ P_{tt}^{n_{_b}-n_{_{bv}}} P_{tb} P_{bb}^{n_{_{bv}}-1}$ & $ P_{tb} P_{bb}^{n_{_b}-1}$ \\ 
\end{tabular}
\label{tab_n_{_{bv}}}
\end{table*}
\begin{figure}[b]
 \centering
\includegraphics*[width=1\columnwidth]{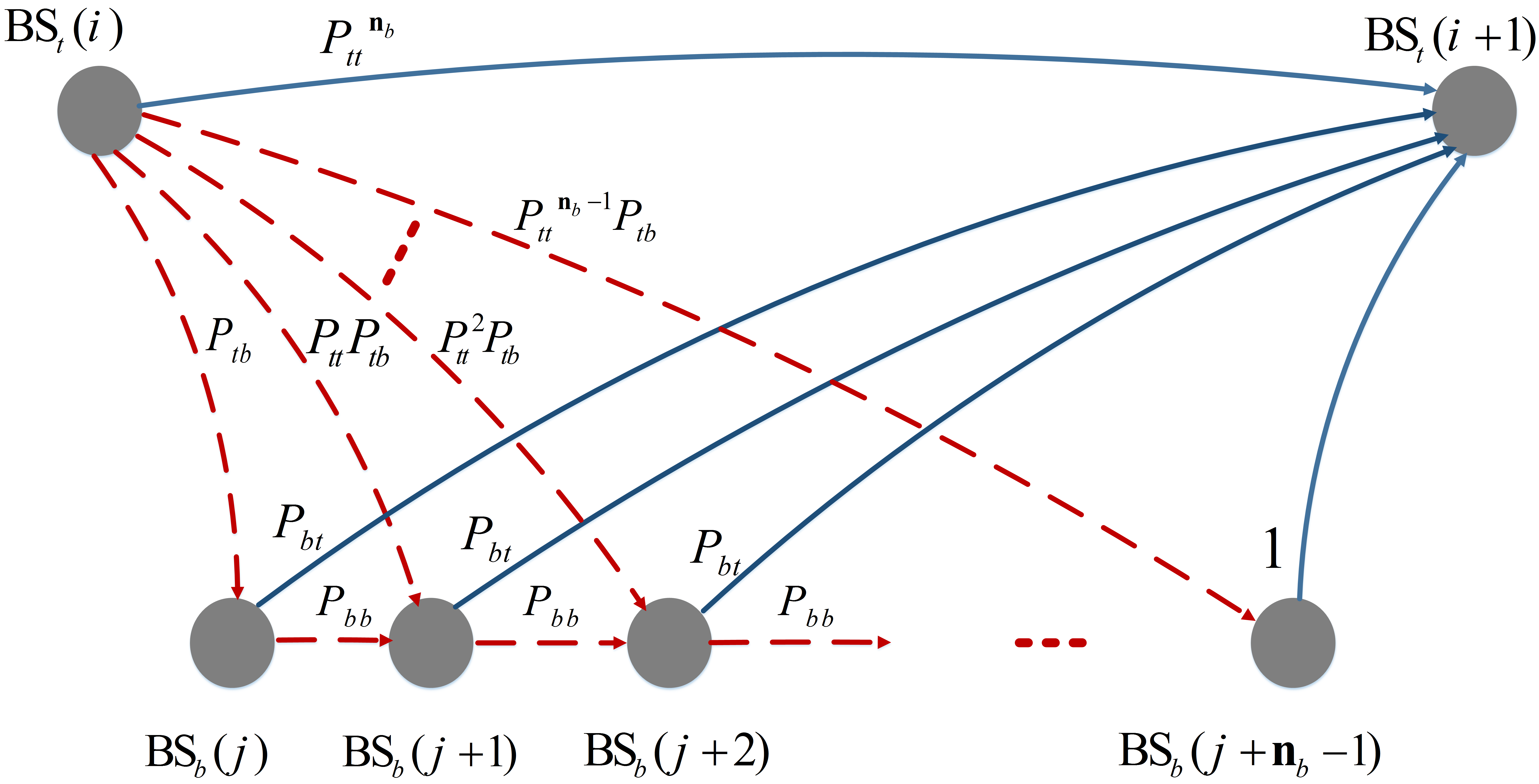}
\caption{The state machine of  the handover events. The solid lines and the dashed lines respectively refer to handover to the BSs at the top side and bottom side of the highway.}
 \label{state_machine}
\end{figure}

Thus,
  \begin{align}
  \label{eq_E_ns}
&  \mathbb{E}_{\mathbf{n}_{_{bv}}}\{\mathrm{NS}_{t,\mathrm{box}} | \mathbf{n}_{_{b}}=n_{_b},\mathbf{d}_{t}=r  \}=\\&\begin{cases}
 \mathrm{N}_{t}(r) & n_{_b}=0 \\ \\
\mathrm{N}_{t}(r)P_{tt}+(\mathbb{E}\{\mathrm{N}_{tb}\}+\mathbb{E}\{\mathrm{N}_{bt}\})P_{tb}  & n_{_b}=1\\ \\
 \Bigg[\mathrm{N}_{t}(r)P_{tt}^{n_{_b}}+
 \Big(\mathbb{E}\{\mathrm{N}_{tb}\} +& n_{_b}\geq 2
\\ \mathbb{E}\{\mathrm{N}_{bt}\}\Big)\Big(\sum\limits_{n=1}^{n_{_b}-1} P_{tt}^{n-1}P_{tb}P_{bt}+P_{tt}^{n_{_b}-1}P_{tb}\Big)
\\  +\sum\limits_{n_{_{bv}}=2}^{n_{_b}-1}\Big(\mathbb{E}\{\mathrm{N}_{tb}\}+
   (n_{_{bv}}-1)\mathbb{E}\{\mathrm{N}_{b}(\mathbf{d}_b)\} +\\  \mathbb{E}\{\mathrm{N}_{bt}\}\Big)\\ \Big( \sum\limits_{n=1}^{n_{_b}-n_{_{bv}}}P_{tt}^{n-1} P_{tb} P_{bb}^{n_{_{bv}}-1}P_{bt}+ P_{tt}^{n_{_b}-n_{_{bv}}} P_{tb} P_{bb}^{n_{_{bv}}-1}\Big) +\\ \Big(\mathbb{E}\{\mathrm{N}_{tb}\} + \\ (n_{_b}-1)\mathbb{E}\{\mathrm{N}_{b}(\mathbf{d}_b)\}+ \mathbb{E}\{\mathrm{N}_{bt}\}\Big)P_{tb}P_{bb}^{n_{_b}-1}\Bigg]  
  \end{cases}\notag
  \end{align}
\\  
Consequently, employing \eqref{eq_Num_tt} and \eqref{eq_E_ns}, we have
\begin{align}
&\mathbb{E}\{\mathrm{NS}_{t,\mathrm{box}}\} \notag \\& \notag
=\int_{0}^{\infty} \lambda_{t,L}\mathrm{N}_{t}(r) e^{-\lambda_{_{t,b,L}}r} dr \\
 \notag & +P_{tt}\int_{0}^{\infty}  \lambda_{t,L} \lambda_{b,L} r
\mathrm{N}_{t}(r)e^{-\lambda_{_{t,b,L}} r}dr 
\\&+(\mathbb{E}\{\mathrm{N}_{tb}\}+\mathbb{E}\{\mathrm{N}_{bt}\})P_{tb}
\int_{0}^{\infty}\lambda_{t,L}  \lambda_{b,L} r 
e^{-\lambda_{_{t,b,L}} r} dr  \notag
\\&
+\sum\limits_{n_{_b}=2}^{\infty}P_{tt}^{n_{_b}} \int_{0}^{\infty}  \lambda_{t,L} (\lambda_{b,L})^{n_{_b}} r^{n_{_b}} 
\mathrm{N}_{t}(r)\dfrac{e^{-\lambda_{_{t,b,L}} r}}{n_{_b}!} dr
\notag \\& \label{eq_E_tot_1}
+\sum\limits_{n_{_b}=2}^{\infty}\Bigg[\Big[ \Big(\mathbb{E}\{\mathrm{N}_{tb}\} + \mathbb{E}\{\mathrm{N}_{bt}\}\Big)\Big(\sum\limits_{n=1}^{n_{_b}-1} P_{tt}^{n-1}P_{tb}P_{bt}+P_{tt}^{n_{_b}-1}P_{tb}\Big)
\\ & +\sum\limits_{n_{_{bv}}=2}^{n_{_b}-1}\Big(\mathbb{E}\{\mathrm{N}_{tb}\}+
   (n_{_{bv}}-1)\mathbb{E}\{\mathrm{N}_{b}(\mathbf{d}_b)\} + \mathbb{E}\{\mathrm{N}_{bt}\}\Big)\notag\\& \Big( \sum\limits_{n=1}^{n_{_b}-n_{_{bv}}}P_{tt}^{n-1} P_{tb} P_{bb}^{n_{_{bv}}-1}P_{bt}+ P_{tt}^{n_{_b}-n_{_{bv}}} P_{tb} P_{bb}^{n_{_{bv}}-1}\Big) +\notag\\& \Big(\mathbb{E}\{\mathrm{N}_{tb}\} + (n_{_b}-1)\mathbb{E}\{\mathrm{N}_{b}(\mathbf{d}_b)\}+ \mathbb{E}\{\mathrm{N}_{bt}\}\Big)P_{tb}P_{bb}^{n_{_b}-1}\Big]\notag\\& \int_{0}^{\infty}\lambda_{t,L}  (\lambda_{b,L})^{n_{_b}} r^{n_{_b}} 
\dfrac{e^{-\lambda_{_{t,b,L}} r}}{n_{_b}!} dr\Bigg]  
\notag
\end{align}
Now it is sufficient that we solve the following integrals:
\begin{align}
&\int_{0}^{\infty} \lambda_{t,L}\mathrm{N}_{t}(r) e^{-\lambda_{_{t,b,L}}r} dr \label{int_1} \\
& 
\int_{0}^{\infty}  \lambda_{t,L} (\lambda_{b,L})^{n_{_b}} r^{n_{_b}} 
\mathrm{N}_{t}(r)\dfrac{e^{-\lambda_{_{t,b,L}} r}}{n_{_b}!} dr  \label{int_2}
\\
& \int_{0}^{\infty}\lambda_{t,L}  (\lambda_{b,L})^{n_{_b}} r^{n_{_b}} 
\dfrac{e^{-\lambda_{_{t,b,L}} r}}{n_{_b}!} dr \label{int_3}
\end{align}
1) Evaluation of \eqref{int_1}:
\begin{align}
&\int_{0}^{\infty} \lambda_{t,L}\mathrm{N}_{t}(r) e^{-\lambda_{_{t,b,L}}r} dr= \notag \\
&=\sum\limits_{k=1}^{\frac{N_c}{4}-1}k\int_{2W_{t,l} a_{_{k-1}}}^{2W_{t,l} a_{_{k}}}\lambda_{t,L} e^{-\lambda_{_{t,b,L}}r}
dr\notag\\&+\frac{N_c}{4}\int_{2W_{t,l} a_{_{\frac{N_c}{4}-1}}}^{\infty}\lambda_{t,L} e^{-\lambda_{_{t,b,L}}r} \notag
dr\\
&=\frac{\lambda_{t,L}}{\lambda_{_{t,b,L}}}\sum\limits_{k=0}^{\frac{N_c}{4}-1} e^{-2W_{t,l} \lambda_{_{t,b,L}} a_{_k}} \label{solve_int_1}
\end{align}
\\
2) Evaluation of \eqref{int_2}:
\begin{align}
\notag&\int_{0}^{\infty}  \lambda_{t,L} (\lambda_{b,L})^{n_{_b}} r^{n_{_b}} 
\mathrm{N}_{t}(r)\dfrac{e^{-\lambda_{_{t,b,L}} r}}{n_{_b}!} dr =
\\
 \notag& =\frac{(\lambda_{b,L})^{n_{_b}}\lambda_{t,L}}{n_{_b}!}\Big(\sum\limits_{k=1}^{\frac{N_c}{4}-1} k\int_{2W_{t,l} a_{_{k-1}}}^{2W_{t,l} a_{_k}} r^{n_{_b}} e^{-\lambda_{_{t,b,L}} r} dr
 \\ \notag &+
\int_{2W_{t,l} a_{_{\frac{N_c}{4}-1}}}^{\infty} \frac{N_c}{4} e^{-\lambda_{_{t,b,L}} r} dr\Big)=
\frac{(\lambda_{b,L})^{n_{_b}}\lambda_{t,L}}{n_{_b}!}\Big(\sum\limits_{k=1}^{\frac{N_c}{4}-1} k 
\\&  \Big[\sum\limits_{m=1}^{n_{_b}+1}\dfrac{n_{_b}!r^{(n_{_b}-m+1)}(-1)^{(m-1)}e^{-\lambda_{_{t,b,L}}  r}}{(n_{_b}-m+1)!}(\dfrac{-1}{\lambda_{_{t,b,L}}} )^m\Big]_{2W_{t,l} a_{_{k}}}^{2W_{t,l} a_{_{k-1}}} \notag\\& +
\frac{N_c}{4}\Big[\sum\limits_{m=1}^{n_{_b}+1}\dfrac{n_{_b}!r^{(n_{_b}-m+1)}(-1)^{(m-1)}e^{-\lambda_{_{t,b,L}}  r}}{(n_{_b}-m+1)!}\label{solve2}\\ &(\dfrac{-1}{\lambda_{_{t,b,L}}} )^m\Big]_{2W_{t,l} a_{_{\frac{N_c}{4}-1}}}^{\infty} \Big)\notag \\& 
=\lambda_{t,L}\sum\limits_{m=1}^{n_{_b}+1}\sum\limits_{k=0}^{\frac{N_c}{4}-1} \dfrac{(\lambda_{b,L})^{n_{_b}}(2W_{t,l}a_{_k})^{n_{_b}-m+1} }{(n_{_b}-m+1)!(\lambda_{_{t,b,L}})^m}e^{-2W_{t,l} \lambda_{_{t,b,L}} a_{_k}}\notag
\end{align}
3) Evaluation of \eqref{int_3}:
\begin{align}
& \int_{0}^{\infty}\lambda_{t,L}  (\lambda_{b,L})^{n_{_b}} r^{n_{_b}} 
\dfrac{e^{-\lambda_{_{t,b,L}} r}}{n_{_b}!} dr\notag
\\&=
\frac{\lambda_{t,L}(\lambda_{b,L})^{n_{_b}}}{n_{_b}!}\int_{0}^{\infty} r^{n_{_b}}e^{-\lambda_{_{t,b,L}} r}dr \notag
\\&=
\frac{\lambda_{t,L}(\lambda_{b,L})^{n_{_b}}}{n_{_b}!}\Big[(-1)^{n_{_b}}n_{_b}!\big(\frac{-1}{\lambda_{_{t,b,L}}}\big)^{n_{_b}+1}e^{-\lambda_{_{t,b,L}} r} \Big]_{0}^{\infty}\notag
\\&=-\dfrac{\lambda_{t,L}(\lambda_{b,L})^{n_{_b}}e^{-2\lambda_{_{t,b,L}}}}{(\lambda_{_{t,b,L}})^{n_{_b}+1}}\Big|_{0}^{\infty}
=\dfrac{\lambda_{t,L}(\lambda_{b,L})^{n_{_b}}}{(\lambda_{_{t,b,L}})^{n_{_b}+1}}\label{solve3}
\end{align}

Now, substituting \eqref{solve_int_1}, \eqref{solve2} and \eqref{solve3} in \eqref{eq_E_tot_1}, equation \eqref{eq1_th_beam_switch}
 would be resulted.

In order to obtain $\mathbb{E}\{\mathrm{NH}_{t,\mathrm{box}}\}$ it is sufficient that in \eqref{eq_E_tot_1} we substitute $\mathrm{N}_{tb}$, $\mathrm{N}_{bt}$, $\mathrm{N}_{b}(\mathbf{d}_b)$ and $\mathrm{N}_{t}(\mathbf{d}_t)$
 by 1 then using \eqref{int_3}, equation \eqref{eq2_th_beam_switch}  would be resulted.
\end{document}